# Harnessing the Duality of Magnetism and Conductivity: A Review of Oxide based Dilute Magnetic Semiconductors

Pankaj Bhardwaj[1], Jarnail Singh[2*], Vikram Verma[1] and Ravi Kumar[1]

[1]Department of Material Science and Engineering, National Institute of Technology Hamirpur-177005

[2]Department of Mechanical Engineering, University Centre for Research and Development, Chandigarh University, Mohali 140413, Punjab, India

[*]*E-mail address:* *jarnailnith@gmail.com*

***ORCID ID:*** *0000-0001-6569-9359*

**Abstract**

Over the last two decades, the new branch of spintronics, i.e., semiconductor spintronics, has gained more attention because it integrates the properties of conventional semiconductors, such as optical bandgap and charge carriers, helpful for processing and computing pieces of information combined with magnets for data storage applications in a single device. Likewise, substituting transition metal (TM) ions to induce magneti*sm* into semiconductors or oxides creates dilute magnetic semiconductors (DMSs) or oxides (DMOs) with high electronic, photonic, and magnetic functionality. This review article discusse*d* the historical outline of magnetic semiconductors with their origin and mechanism. It also includes a concise overview of various DMO systems based on their conductivity (*p*-type and *n*-type) to elucidate the synthesis, origin, and control mechanisms and further evoke the prepared spintronics devices. The occurrence of RTFM with transparency and conductivity can be helpful in spintronics device fabrications, which was assumed to be governed by the formation of intrinsic defects, charge carriers, morphology, and the induced exchange interactions between ions. The DMOs-based spintronics devices, such as magneto-optical devices, transparent ferromagnets, and spin-based solar cells, exploit both semiconducting and magnetic properties, which have also been discussed in this review article with outlook and perspectives.

## 1. Background and Introduction

In 1959, Richard Feynman delivered a lecture at the American Physical Society annual meeting entitled "*There plenty of room at the bottom.*" He coined the idea of controlling atoms and molecules in a confined space [1]. Fifteen years later, in 1979, Norio Taniguchi was the first to use the word "nanotechnology" and stated a definition that "Nanotechnology mainly comprises the processing of separation, consolidation, and deformation of materials by one atom or one molecule." The field attracted significant interest and attention because of changes in the physical properties of materials from bulk counterpart to nanoscale level because of the heightening in surface to the volume ratio. This confinement of atoms in smaller dimensions led to changes in the nanostructure, which altered the material's structural, optical, electrical, and magnetic properties, which were generally not explained by classical physics.

Nanotechnology is the blend of science, engineering, and technology that involves altering nanoscale properties to develop efficient materials or create new technology for humanity[2].

One of the fundamental driving forces of nanoscience or nanotechnology is the electronics industry, as the size of devices decreases rapidly with high functionality. Over the past decades, semiconductors reformed the electronics industry from the discovery of the silicon transistor, which led to the development of integrated circuits in 1958 by Jack Kilby and got a Nobel prize in physics in the year 2000 [3]. Years later, in 1965, Gordon Moore postulated a trend that the number of transistors in dense integrated circuits doubled every two years, later termed as Moore’s law. Moore’s law predictions were economic as the multiple transistor or devices in a single chip. However, the pace of Moore’s law slipped in the previous two decades as it became harder to make smaller transistors. The scaling and accommodation of transistors in small chips cause leakage, resulting in more energy dissipation and high-power consumption, which reduces the chip stability and the output of electronic devices [4]. Thus, the evolution of charge-based conventional electronic devices is stagnated, which demands a newer approach for upcoming electronic devices for further miniaturization and to keep going with the prediction of Moore’s law, leading to the next generation of promising alternative information and electronic technology known as Spintronics [5].

Spintronics is an adjective for SPIN TRansfer electrONICS, introduced in 1996 by DARPA to develop the non-volatile MRAM device and magnetic sensors for specialized applications [8]. Spintronics is an emergent technology to solve the problem that arises from conventional electronics. Compared to traditional electronics, which use the degree of electron charge to process information, spintronics exploits both degree of charge and spins of electrons. In contrast, spin is known as the origin of magnetism, where they carry and process information in spin up (1 bit ↑) and spin down (0 bit↓) [6,7]. The first breakthrough in spintronics was the discovery of giant magnetoresistance (GMR) in 1988, independently by Albert Fert and Peter Grünberg, who later received the Noble prize in physics in 2007. They influence the electron spin by controlling electric current by switching between the nonmagnetic and magnetic layers, leading to a rise in the density of stored information and magnetic sensors used in hard disk drives as consumer electronics [9]. Spin in the semiconductors opens a pathway in conventional electronics towards integration of processing and storage of information, which can be used in microelectronics and storage devices. Spintronics improves a single chip's speed, functionality, power consumption, and storage, making the device smaller and faster than conventional electronics. This opened the path for researchers to manipulate spin for further applications by spin polarizing nonmagnetic materials by a different technique, such as spin injection from FM material, magnetic field, electric field, thermal gradient, and photoexcitation [6,7], as shown in Fig. 1(a). It leads to the formation of different potential spintronics devices as illustrated in Fig. 1(b).

(a)

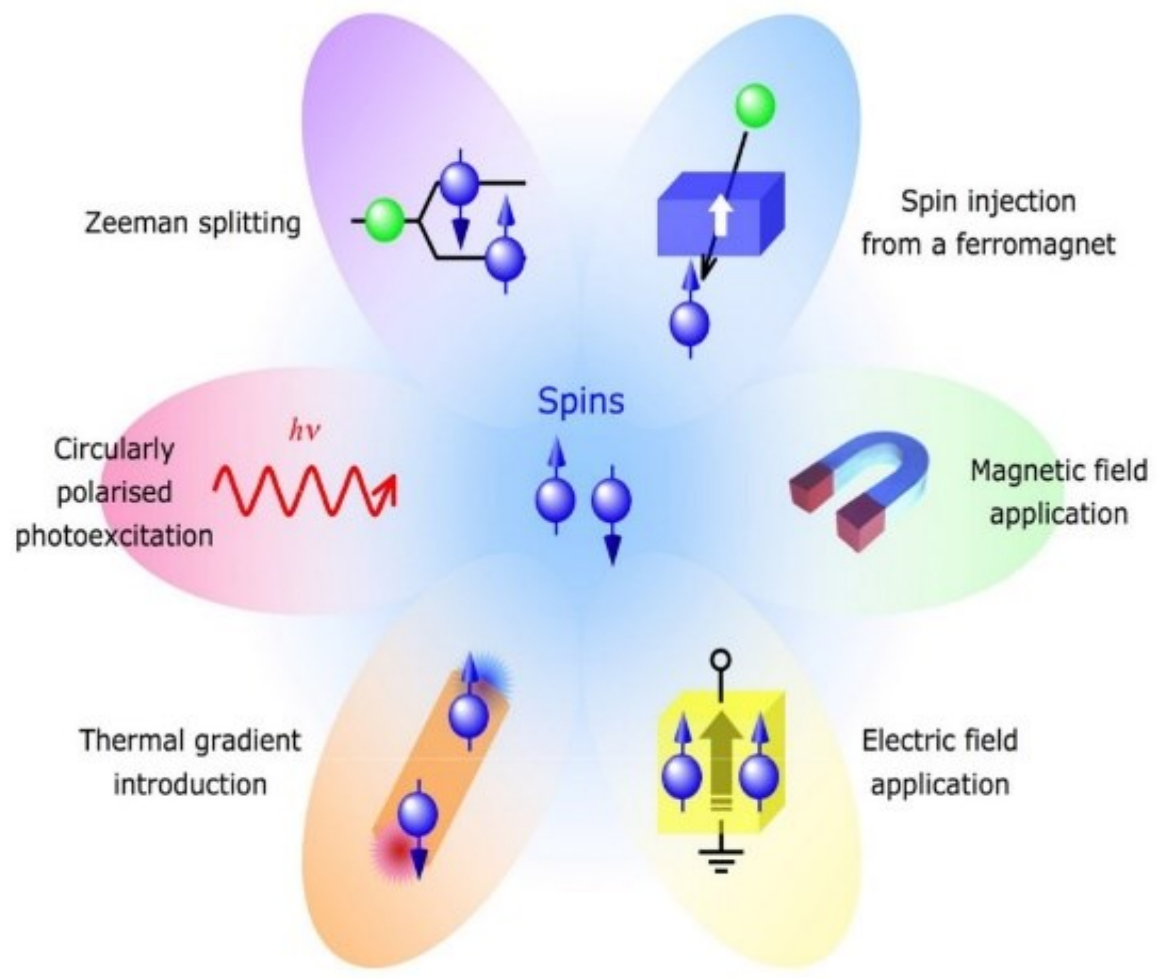

(b)

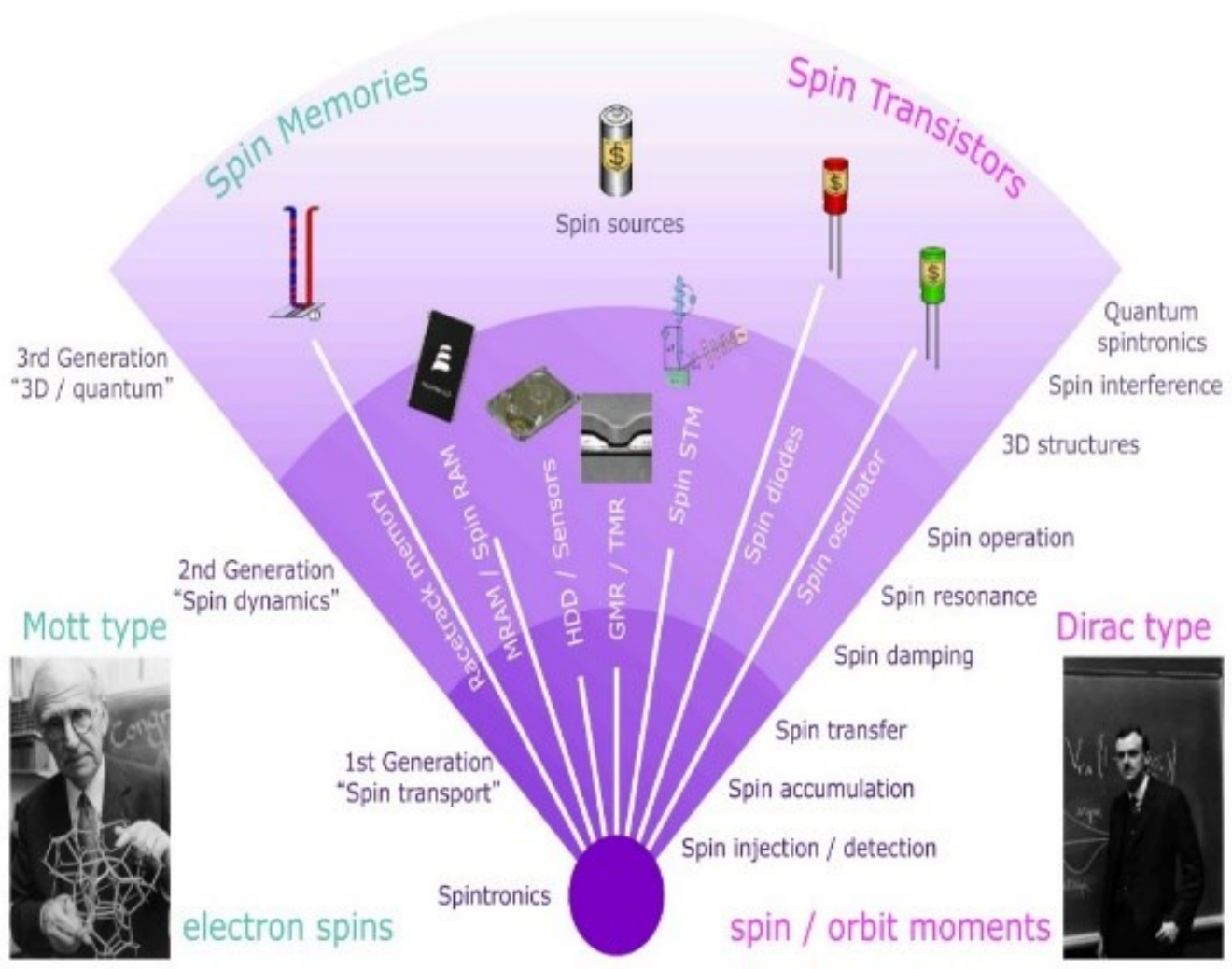

**Figure 1.** (a) List of techniques to generate spin-polarized electron in non-magnetic Material. (b) List of spintronics devices originated from spin generation [6,7].

Based on the spin generation and according to the material properties such as electronics, semiconducting and magnetic properties, the types of spintronics material are classified into three categories as depicted in Fig. 2 [8]. The first is magnetic metal, the traditional and oldest spintronics used to make magnetic tunnel junctions (MTJ) and spin valves. These magnetic metals are further categorized as ferromagnetic metal and half metal, the ferromagnetic transition metal ions and their alloys are used, such as Co, Fe, and Ni metals, as they are cheap and easy to handle, but they supply partial spin-polarized carriers due to low degree of spin polarization. Another is a half-metal called Half-metallic ferromagnets (HMF), discovered by Groot and Mueller in 1983. Their characteristics lie in 100% spin polarization, used for spin injection and generation [9]. The property of spin polarization causes magnetic metal to be generally used as the spin source and drain in spintronics devices [10].

(a)

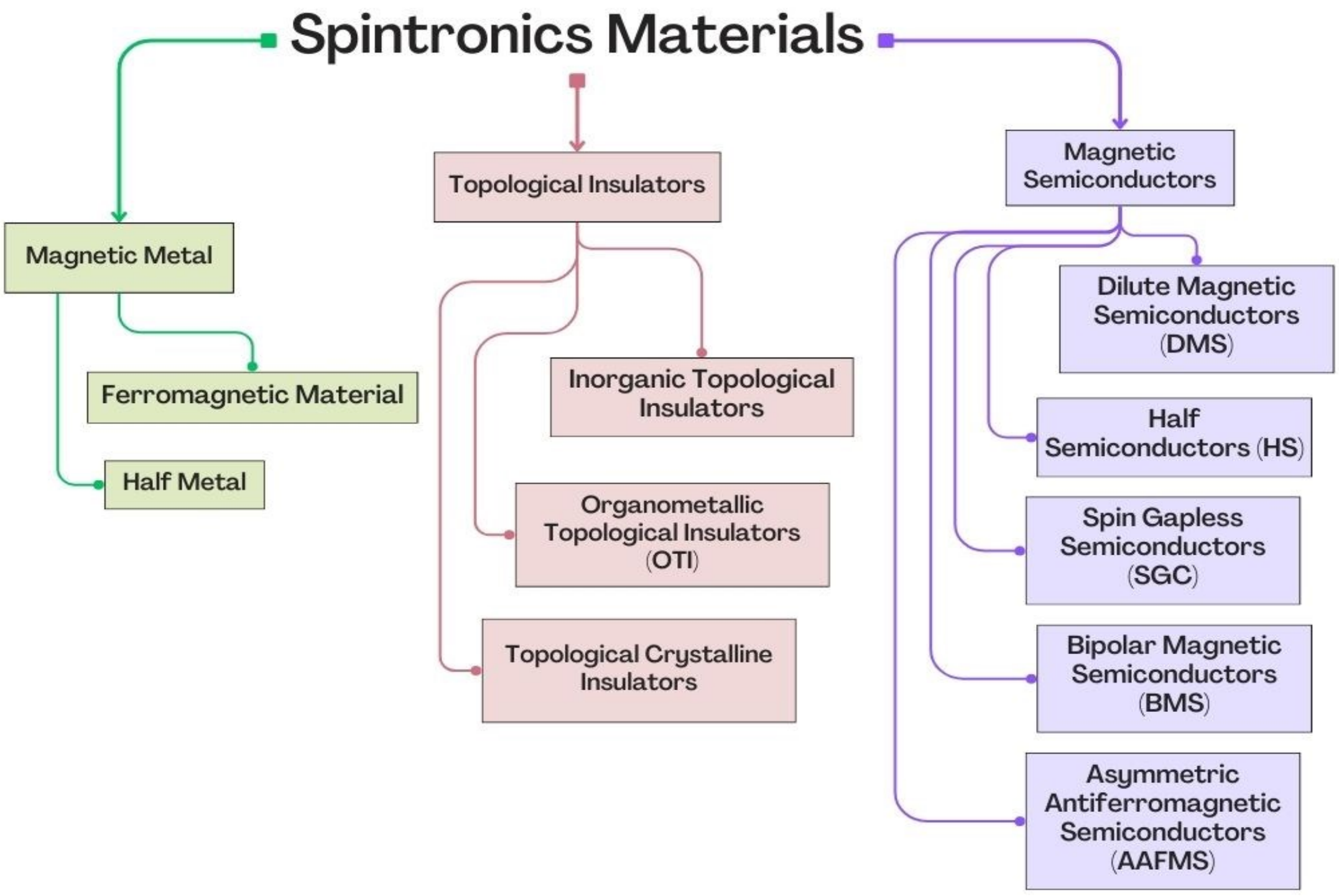

(b)

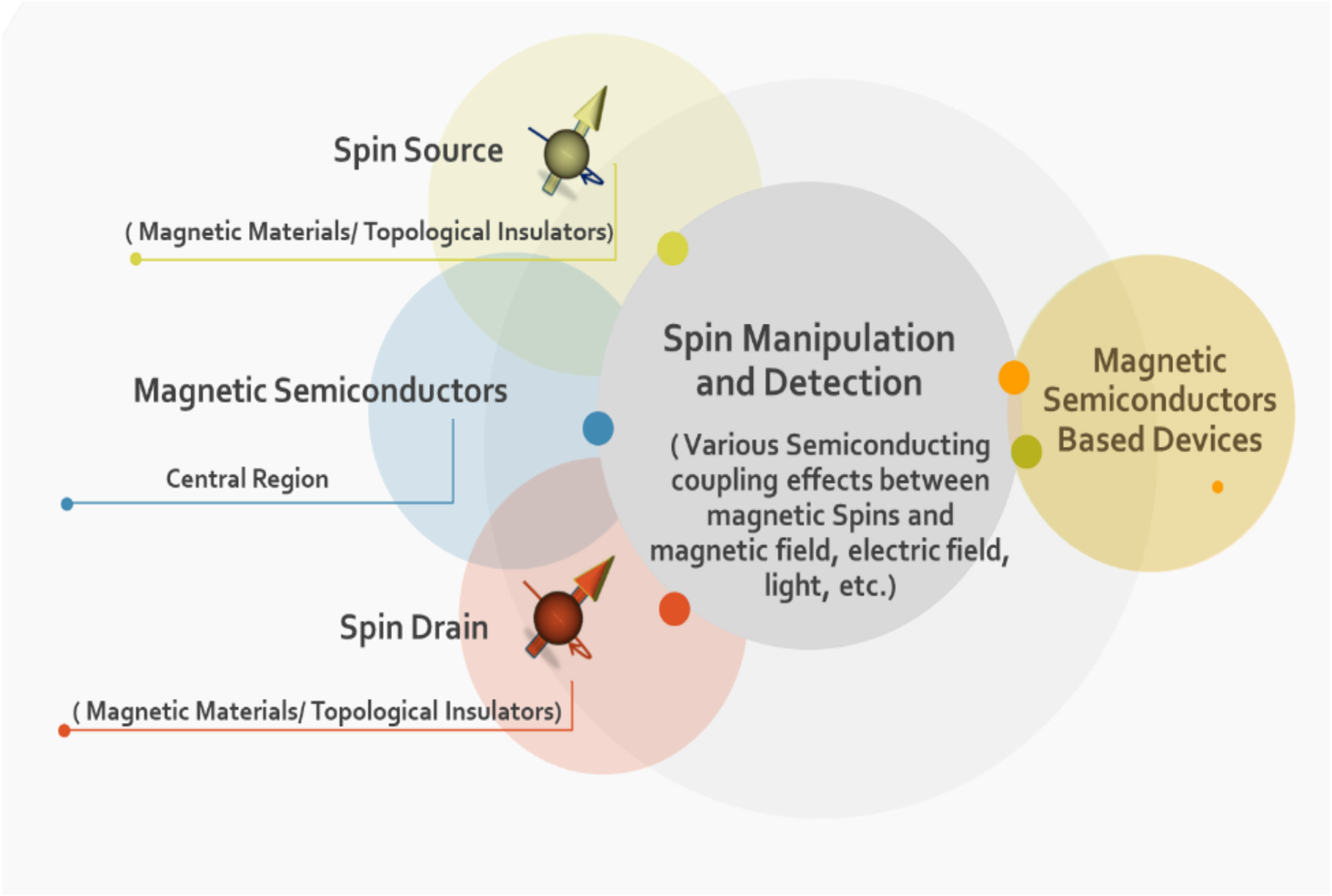

**Figure 2. (a)** Categories of spintronics materials. **(b)** Spins interplay in spintronics material.

Another category of spintronics devices is Topological insulators (TI), insulating in bulk and metallic on the surface to propagate spin up and spin down in opposite directions. These properties make TI ideal for spintronics devices, as they are independent on the current for spin generation and transport. However, they still have a long way to go because of non-functionality at room temperature and are still limited to low temperature [11].

## 2. Magnetic Semiconductors

The third category of spintronics is magnetic semiconductors, which combine the properties of the semiconducting and magnetic properties used on spintronics devices.

Magnetic semiconductors exploit magnetic and well-established semiconductor technology, which can be employed for spin sources, generation, transport, and drain applications. Fig. 3. shows the schematic representation of magnetic semiconductors.

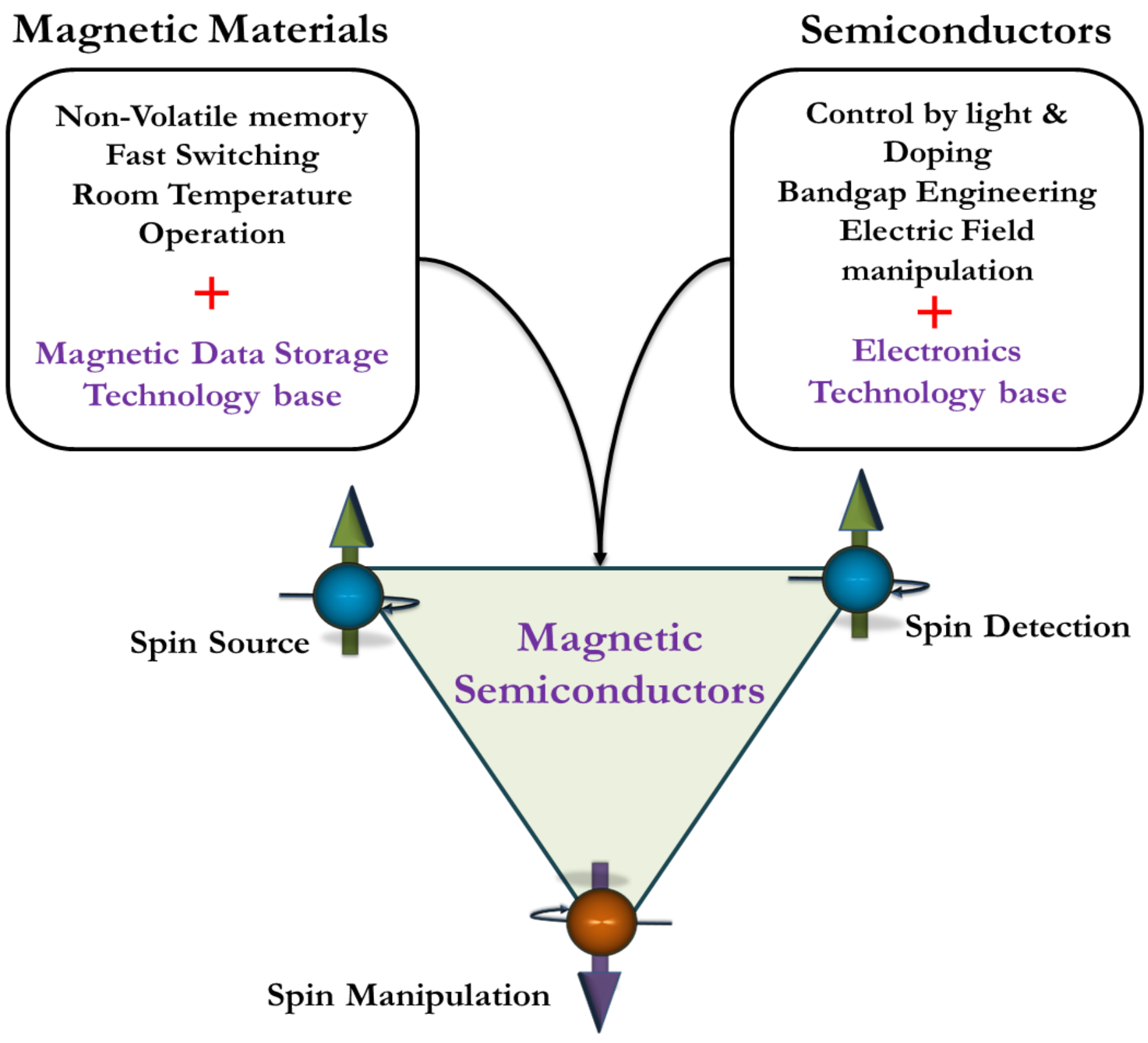

**Figure 3.** The exclusive feature of magnetic semiconductors is the combination of ferromagnetic and semiconductor for the creation, manipulation of spins.

In general, magnetic semiconductors are classified as intrinsic and dilute. Intrinsic magnetic semiconductors employ magnetic and semiconducting properties as inherent characteristics, such as semiconducting spinels ($CdCr_2Se_4$ and $CdCr_2S_4$) and chalcogenides (EuO). The difficulty in material preparation and low curie temperature limits the functionality at room temperature and is unsuitable for spintronics applications. Another one is Dilute magnetic semiconductors (DMS), where a nonmagnetic semiconductor is injected with a small magnetic ions impurity to make them magnetic by exploiting both degrees of charge and spin of electrons [5].

## 2.1 History of Magnetic Semiconductors

Introduction of magnetic semiconductors started early in 1960, with the discovery of chalcogenides and spinels having semiconducting and magnetic properties such as $CrBr_3$, EuX (O, S, Se, Te) and $CdCr_2S_4$ [12]. However, because of the difficulty of growing high-quality thin films and the complicated synthesis process, restrict the curie temperature at most 130 K, which lacks suitability at room temperature and does not lead to any spintronics applications. The commercialization of molecule beam epitaxy (MBE) in the early 1980s opened the path for researchers and scientists to develop or grow high-quality materials and heterostructure thin films [13]. MBE and others thin film depositions techniques open the pathways to *develop* and synthesize semi-magnetic and dilute magnetic semiconductors with 3d transition metal ions doping in II-VI, IV-VI, and III-V hosts. Fig. 4. summarizes the historical overview of magnetic semiconductors.

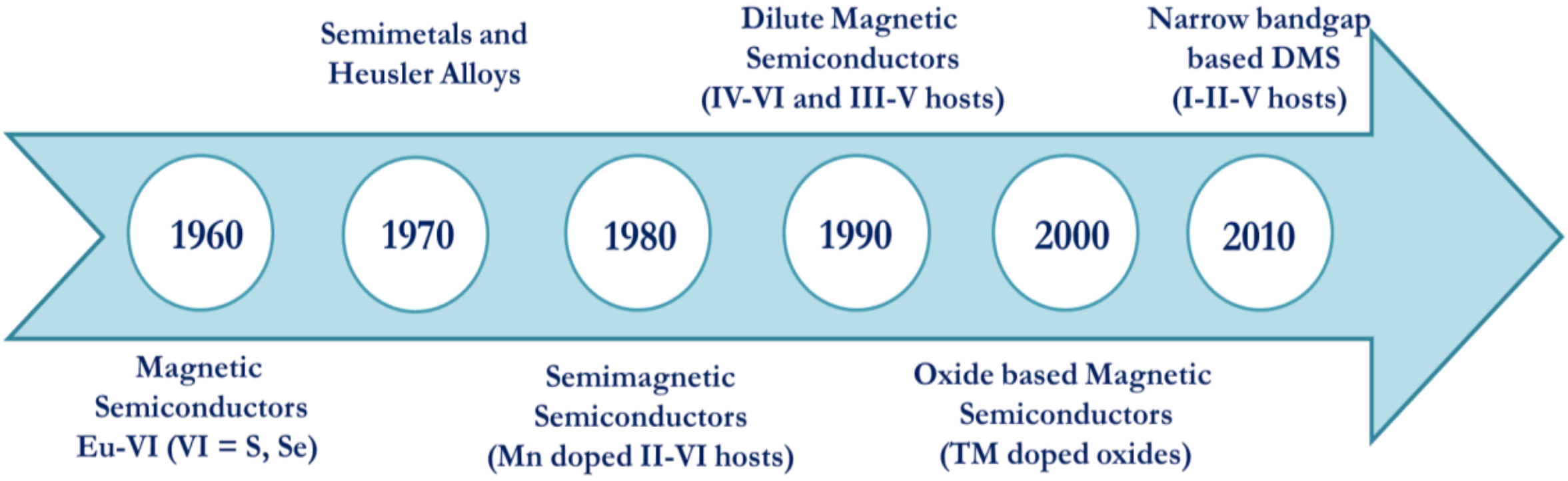

**Figure 4.** Historical outline of Magnetic Semiconductors.

## 2.2 Diluted Magnetic Semiconductors (DMSs)

The small concentration of transition metal ions substituted in a semiconductor compound is such that it does not alter the characteristics of native semiconductors while improving the optical, electrical, and magnetic properties, beneficial for functional spintronics devices at room temperatures defined as dilute magnetic semiconductors (DMSs). The dependence of band structure and the orientation of their electronic spin explained the phenomenon of magnetic and nonmagnetic semiconductors, as the spin-polarized semiconductor bands with non-equilibrium orientations confirm the existence of magnetic semiconductors [14], illustrated in Fig. 5.

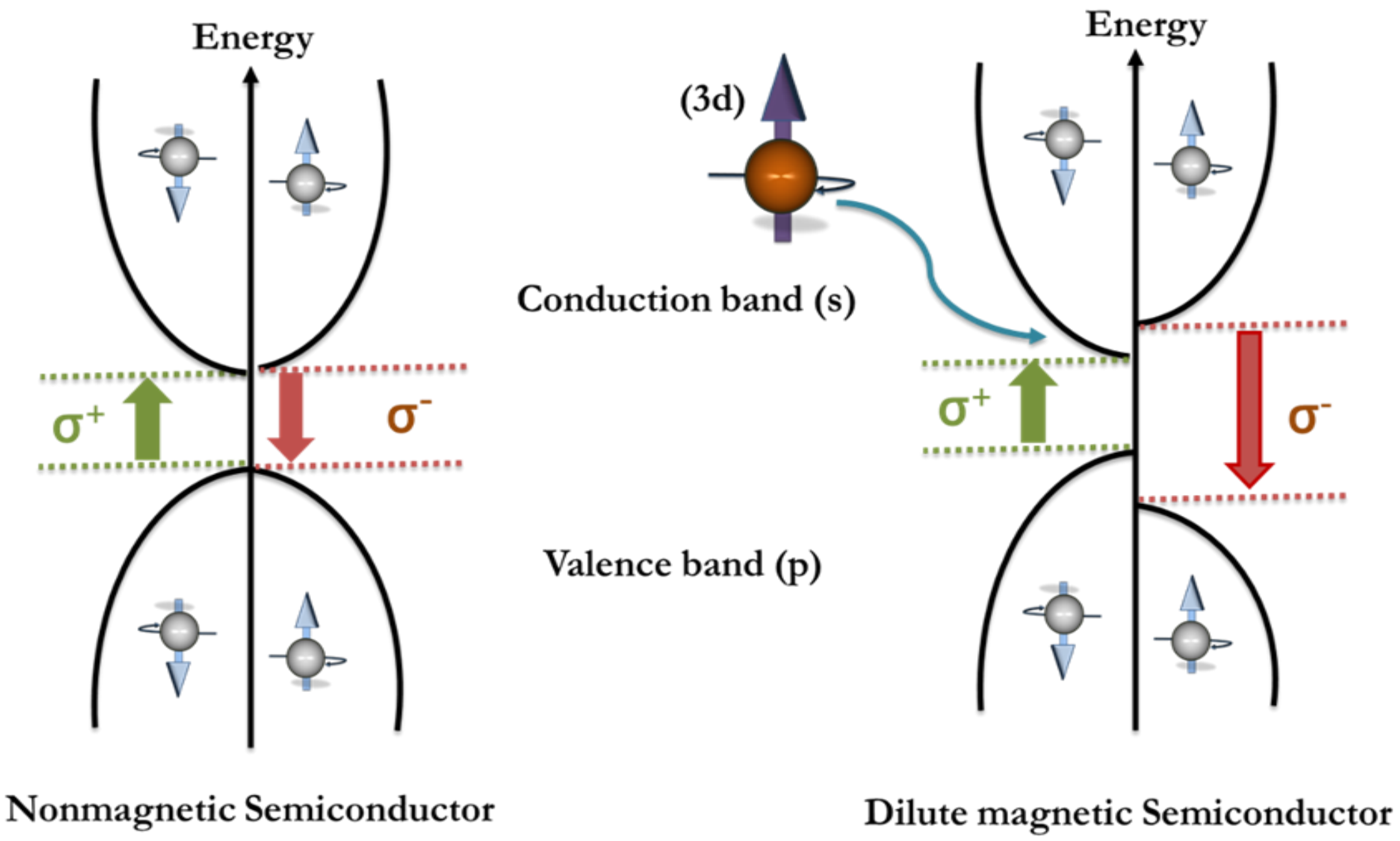

**Figure 5.** Schematic illustration of change in band structure in nonmagnetic semiconductor (**Left**) and dilute magnetic semiconductor (**Right**) with the influence of 3d electrons.

In nonmagnetic semiconductors, the electronic energy of different spin reorientations compensates for each other; thus, it's impossible to distinguish between spin up and spin down as they cancel each other, and only the current of charged particles exists. While in a magnetic semiconductor, the 3d electron of magnetic ions polarized the spins of conduction (s) and valence (p) band electrons, resulting in uneven splitting depending upon the direction of spin, causing the origin of magnetism. The central core of dilute magnetic semiconductors is the observation of magnetic properties from the spin-polarized electron supported by the 3d metal ions and semiconductor properties supported by s and p electrons [14,15].

The first breakthrough in DMS was in the year 1992 when Munekata et al. deposited Mn-substituted *i*ndium arsenide *(*InAs*)* semiconductor via MBE technique and achieved room temperature ferromagnetism (RTFM) with *n*-type conductivity and reduced band gap because of the occurrence of MnAs clusters [16]. Afterwards, in 1992, Ohno et al. deposited Mn (x = 0.015 ~ 0.017) doped GaAs thin films using MBE and got enhanced solubility of Mn without secondary impurities and observed curie temperature ($T_c$) varied as a function of doping concentration [17]. They found that the Mn ions instigate magnetic moments from unpaired electrons and observed *p*-type conductivity having a ferromagnetic order at 7.5 K because of the formation of bound magnetic polarons (BMP) [18]. To enhance the capability of semiconductors, the development of new DMSs materials explored by various researchers such as (Pb,Eu)Te and (Zn,Cr)Se [19], (Cd,Mn)$GeP_2$ and chalcogenides [20] but they all lack room temperature functionality. Thus, developing magnetic semiconductors with the appreciable magnetic moment and semiconducting properties at room temperature became quite challenging in semiconductor spintronics.

## 3. Dilute Magnetic Oxides (DMOs)

In the Years 2000s, Dietl et al. coined a theoretical model that predicted with doping of $Mn^{2+}$ ions (x = 0.05) doping in a *p*-type semiconductors and ZnO having carrier concentration (holes) of 3.5 $\times 10^{20}$ $cm^{-3}$ results in wide optical band gap and high curie temperature, which were explained through indirect hole mediated ferromagnetic super-exchange interaction [21], as shown in Fig.6(a).

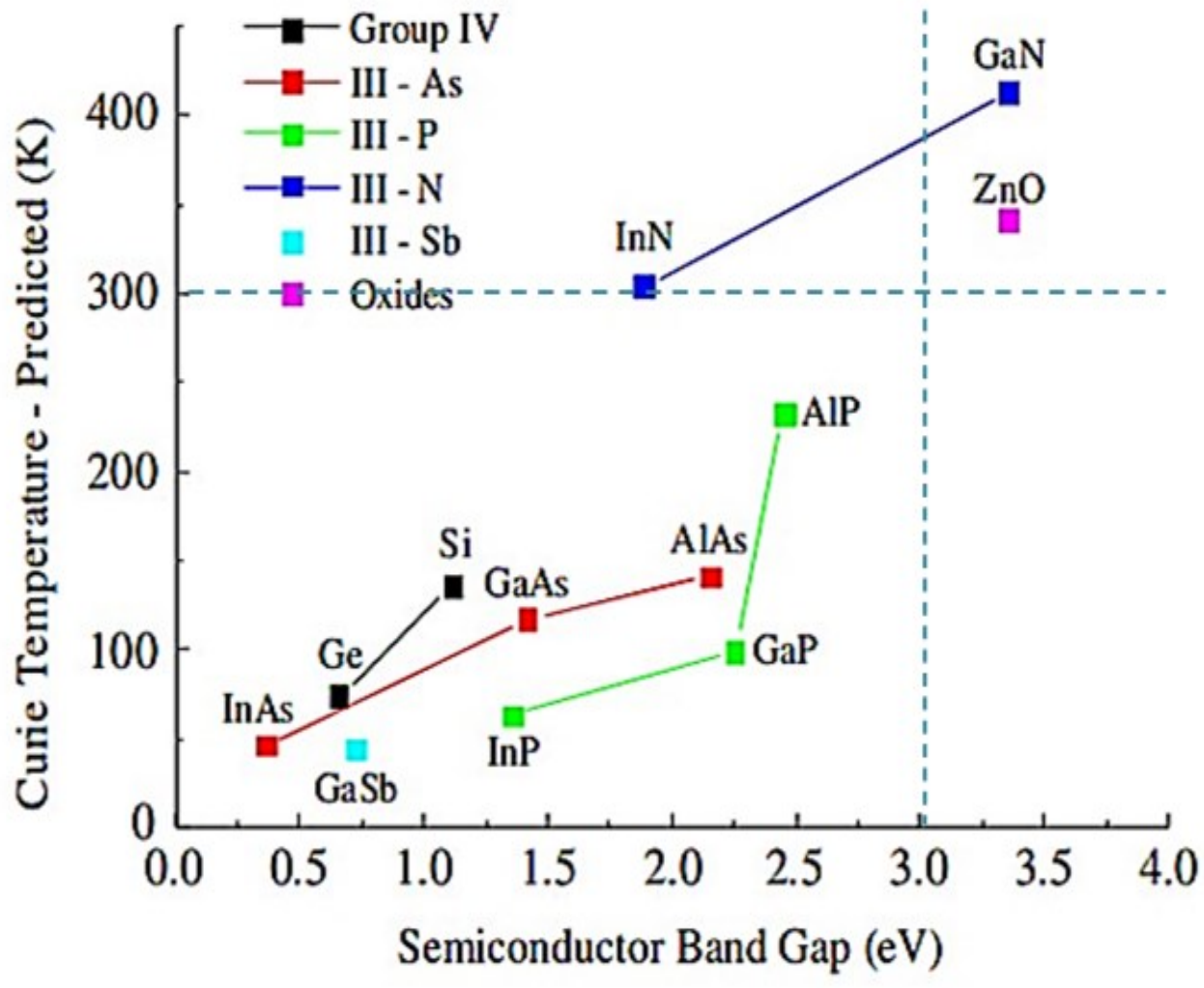

**Figure 6 (a).** Computed values of Curie temperature and Bandgap with 5 % Mn doping in p-type semiconductors and oxide.

Fig. 6(a) trends show that earlier discovered DMS materials with Mn-substitution acquired low optical band gap and curie temperature, which restricts many practical application*s* related to magnetism and semiconducting properties. A year later, in 2001, Zajac et al. grew Mn (x = 0.005) doped GaN microcrystalline sample using an ammonothermal method and observed paramagnetic contribution at room temperature and concluded that observed RTFM in theoretical studies might be because of *p*-type conductivity [22]. The brief trend from DMSs to DMOs and historical overview were found using the research database set by placing Dietl et al. as the origin point using connected papers, as shown in Fig.6 (b).

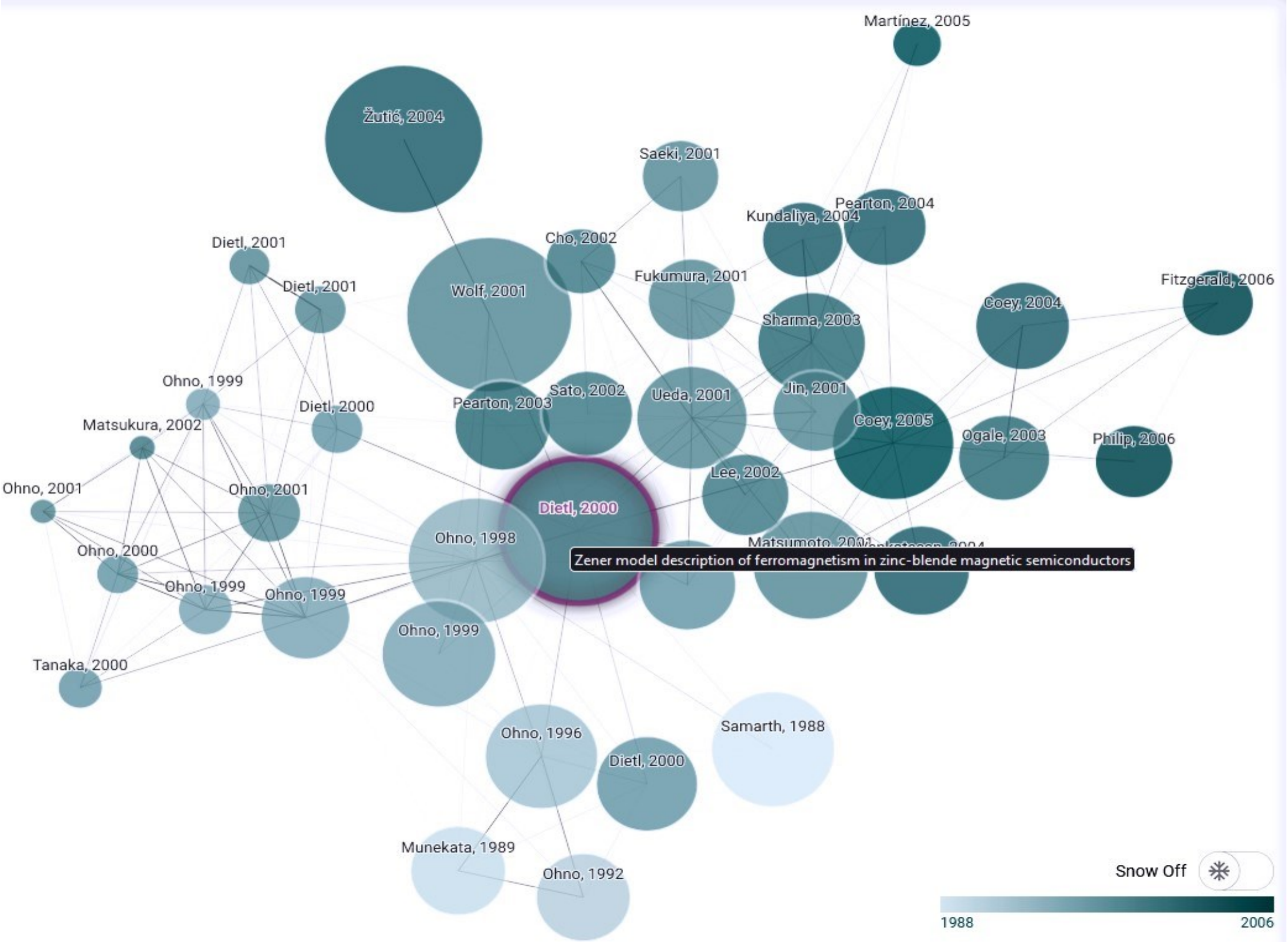

**Figure 6 (b).** The historical outline from Diet et al. (2000) to DMOs [(Matsumoto et al. (2001), Ueda et al. (2001)] plotted using *www.connectedpapers.com.*

Following the year in 2001, the milestone in oxide-based DMSs was came into existence when Matsumoto et al. deposited the Co-doped (x = 0.07) anatase $TiO_2$ thin films using MBE and achieved a wide optical band gap with high transparency and $T_c$ above 400 K with appreciable *n*-type conductivity due to Co cluster [23]. Incorporating TM ions into an oxide material opens the path for better semiconducting properties, with a higher magnetic contribution at room temperature termed *as* oxide based DMS, also known as diluted magnetic oxides (DMOs). The major hindrances in making TM ions diluted III-V and II-VI magnetic semiconductors were the limited solubility of TM ions, complex preparation methods or setups and non-functionality at room temperature. For these reasons, DMOs attain keen interest, which exhibits magnetism from the injection of a few % TM ions and semiconducting properties of native oxides at room temperature.

The general formula of DMOs described by the following formula:

$$(M_{(1-x)}TM_x)O_n \tag{1}$$

where M belongs to the non-metal oxide, TM belongs to substituted TM ions, the concentration of TM ions is depicted by x (≤ 0.1) and n is integer or rational fraction. DMO illustration shown in Fig.7.

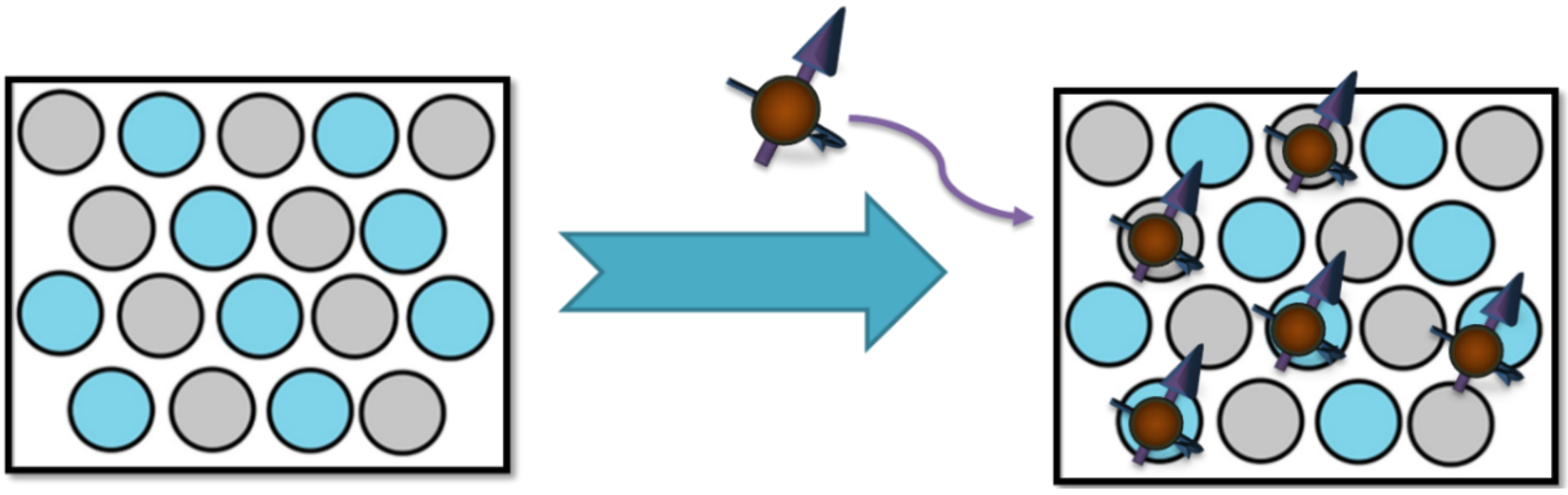

**Figure 7.** Schematic representation of nonmagnetic **(Left)** and dilute magnetic oxides **(Right).**

### 3.1 Source of high $T_c$ in DMOs

The source of high $T_c$ in DMOs contributed by the 3d ions substitution and formation of intrinsic defects or transfer of conduction band electrons, explained by the electronic structure of DMOs as shown in Fig. 8. From Fig. 8(a), it is observed that the location of the 3d band decides the $T_c$ nature of the oxides when the splitting of the impurity band is minor, resulting in low $T_c$. While Fig. 8(b) shows the position of the minority spin 3d band, and Fig. 8(c) shows the position of the majority spin 3d band, which leads to high $T_c$. The electronic band states of DMOs with TM ions substitution led to high $T_c$ only when the empty majority or minority-spin d states are at the Fermi level in the impurity band [24][25].

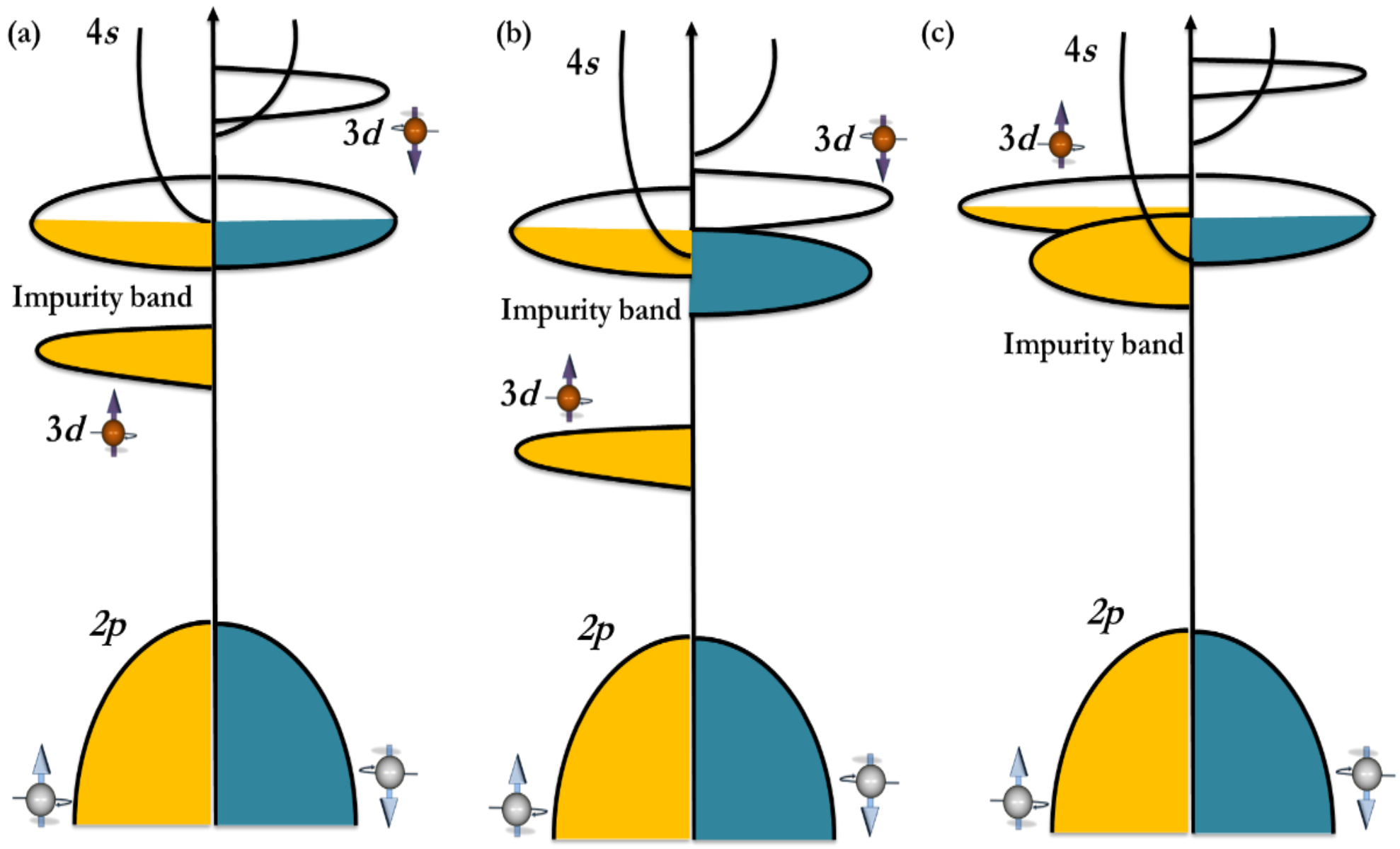

**Figure 8.** Electronic band structure of DMOs with 3d ions substitution.

### 3.2 Why DMOs?

Compared to DMS materials, DMOs materials had more advantages such as [26]:

*a.* Wide optical bandgap suited for optical devices-based application.
*b.* High transparency suited for transparent semiconductor spintronics.
*c.* High *p*-type or *n*-type carrier concentrations suited for localized spin induced magnetism.
*d.* Capability to grow on low temperature and even on any substrate, low deposition cost, ecological safety and high durability make DMOs attractive.

### 3.3 Criteria for DMOs Materials Selection

Significant characteristics are important to obtain viable DMOs material as follows:

*a.* With TM ions substitution in oxide material, DMOs material should show RTFM and had enhanced $T_c$ to above room temperature.
*b.* The parent oxide material should have a strong application base such as wide bandgap, high mobility, transparency, photocatalysis etc.
*c.* The growth of DMOs material should be simple and economically viable.

## 4. Mechanism and Control of DMOs

Among the number of theoretical and experimental models, the origin and mechanism of RTFM in oxides or semiconductors still need to be fully explained and understood. These models have been changed and verified from time to time with subsequent ongoing research. The origin of ferromagnetism in DMOs depends upon the localization or delocalization of electrons and their exchange interaction between the lattices. There are various magnetic interactions in DMOs as follows:

### 4.1 Double Exchange Model

In 1951, Zener postulated a model for FM order based upon the role of band carriers and their indirect exchange interactions between the localized spins and carriers via the intermediary p-orbital anion. This type of interaction is known as a double exchange interaction. Double exchange interaction occurs when the two isolated TM ions contain different numbers of electrons or have a mixed-valence state in the magnetic shell. Their hopping is mediated by the neighboured nonmagnetic atoms causing FM order, as illustrated in Fig. 9.

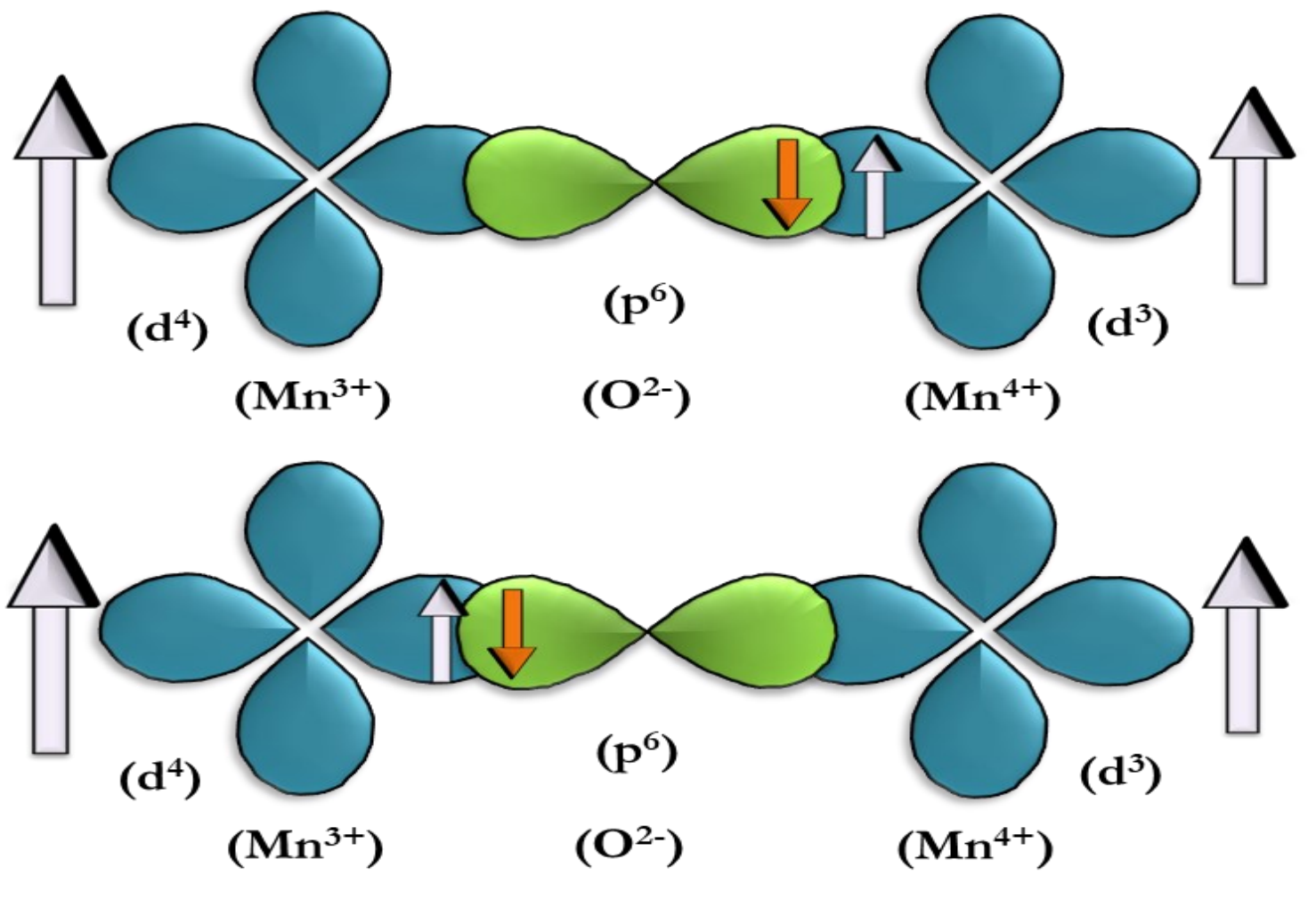

**Figure 9.** Schematic representation of Zener double exchange mechanism.

Zener applied the double exchange mechanism in the manganese perovskite ($LaMnO_3$) to provide the correlation between conductivity and magnetism, which was associated with the occurrence of multiple oxidation states of Mn. He considered the exchange between $Mn^{4+}$ and $Mn^{3+}$ ions via intermediary $O^{2-}$ ion and introduced the notion of indirect transfer of an electron from the $O^{2-}$ anion to the neighbouring $Mn^{4+}$ ions and from the $Mn^{3+}$ ions to the $O^{2-}$ anion. The Hund rules state that the energy of a system is minimized at low temperatures when the configuration magnetic atoms $Mn^{4+}$- $O^{2-}$ - $Mn^{3+}$ and $Mn^{3+}$ - $O^{2-}$ - $Mn^{4+}$ and degenerate, and their spin alignment between the two adjacent cations are parallel to each other, which give FM order [27]. The double exchange model generally favours the interaction between the mixed oxidation state of similar metals and the number of configurations restricted to two [28].

### 4.2 Mean Field Zener Model

With the attention of the Origin of DMS, the proposed Zener model of exchange interactions between localized spins and carriers which drove material to FM, was abandoned because of the non-consideration of electron spin polarization around the localized spins (Friedel oscillations). Later, the Zener model was modified by Dietl et al. [21] for semiconductor materials considering Friedel oscillations and became the mean field Zener model, which was based on the original model of Ruderman, Kittel, Kasuya, and Yosida (RKKY) interactions. Compared to RKKY interaction, the mean-field Zener model considers the anisotropy of magnetic spins and indirect exchange interactions mediated by the localized holes, which cause FM ordering in DMS and DMO materials and further down to road named as indirect hole exchange interactions [26], as shown pictorial depiction in Fig. 10 (a).

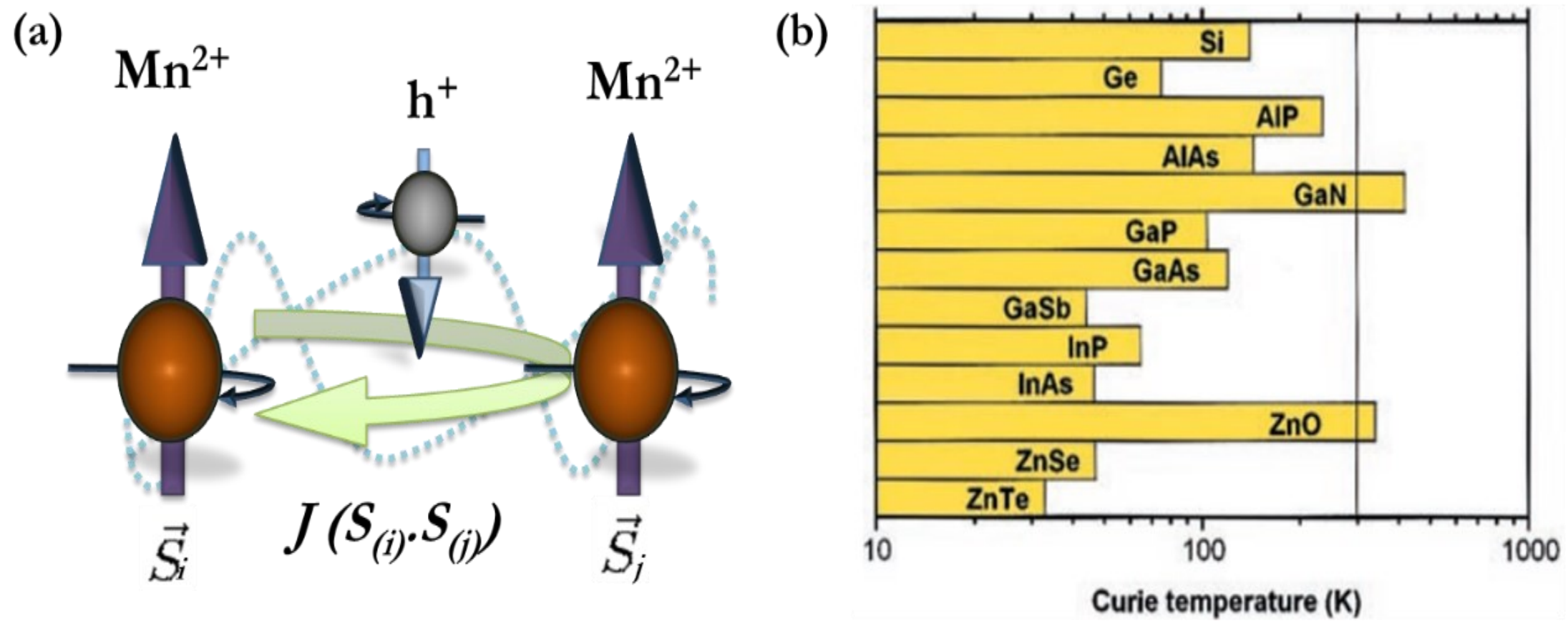

**Figure 10 (a).** Illustration of exchange interaction of Mn ions mediated by delocalized hole **(b)** Theoretical predicted $T_c$ of *p*-type DMS and DMO with 5 % Mn doping [21].

Dietl et al. theoretically predicted that the doping of 5% Mn, which acts as an acceptor in all III-V semiconductors having *p*-type conductivity with a hole concentration of $3.5 \times 10^{20}$ $cm^3$, causes the polarization of localized spin mediated through the hole in a magnetic semiconductor provides FM order and improvement in curie temperature, as shown in Fig. 10 (b). The mean field Zener model also confirms that the higher carrier concentration and magnetic ions concentration in *p*-type ferromagnetic results in the increment of $T_c$ [29].

### 4.3 Ruderman–Kittel–Kasuya–Yosida (RKKY) model

Alike the mean field Zener model, the origin of magnetism in the RKKY model occurs from the indirect magnetic exchange interaction between delocalized band carriers and localized spins in a sea of conduction electrons without considering the magnetic anisotropy of the material. In RKKY interaction, the conduction electron near the magnetic ions gets spin-polarized and acts as a field, which affects the oscillatory spin polarization of the neighbouring atom and causes the nearest atom indirect RKKY exchange interaction. The separation between the magnetic ions ($r_{ij}$), whether parallel (FM) or antiparallel (AFM), decides the magnetic order in the system, as illustrated in Fig. 11.

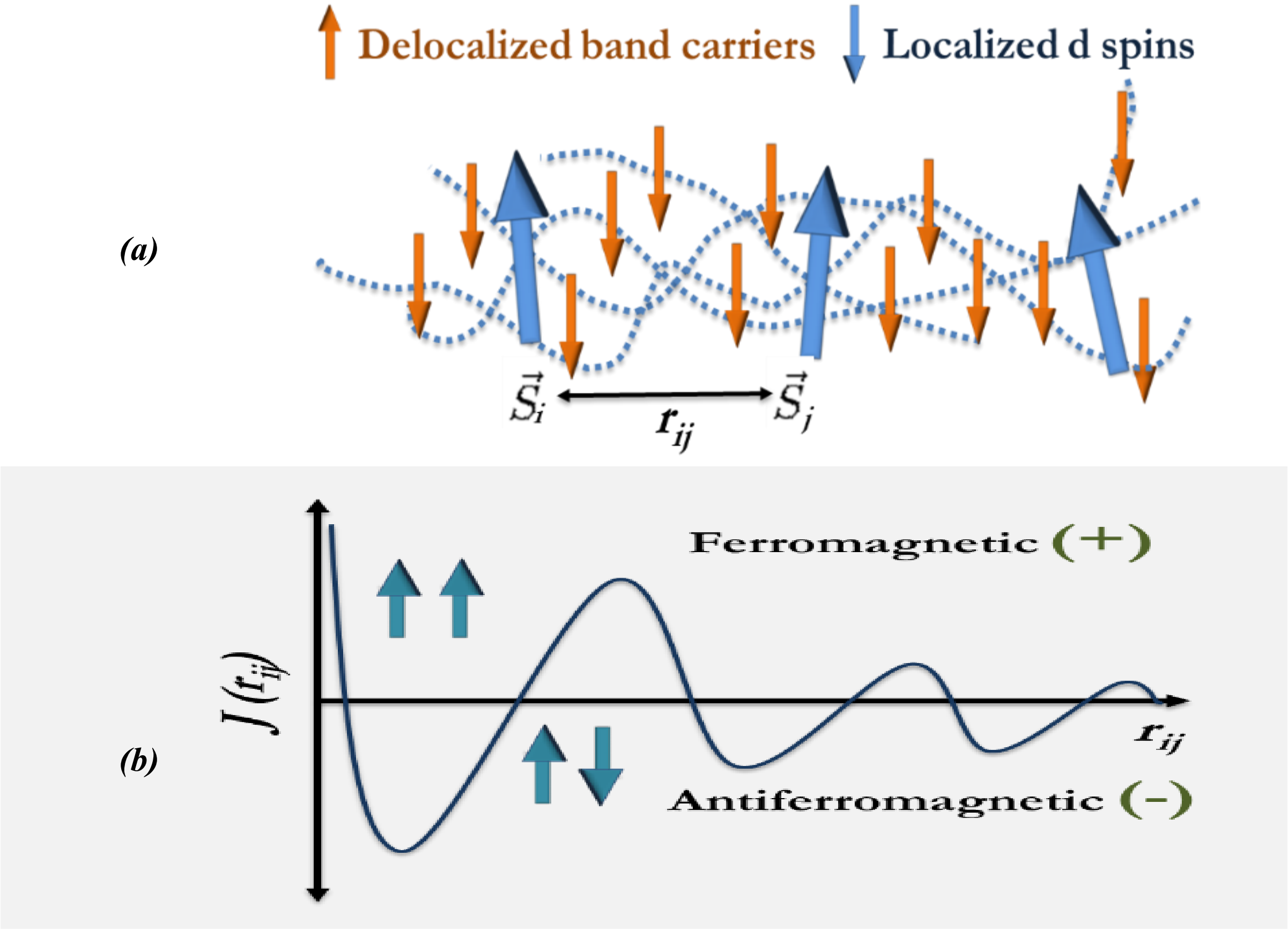

**Figure 11.** (a) Schematic representation of indirect RKKY interactions and (b) Oscillatory nature of RKKY interaction as a function of $r_{ij}$ between the ions.

Thus, RKKY exchange interaction mainly depends upon the carrier concentration of material and primarily exists in the existing metallic systems because of highly delocalized carriers [30].

### 4.4 Super-Exchange Model

The super-exchange phenomenon was first invented by Kramers et al. in 1934 [31], and Anderson developed a further theory in 1950 [32]. Usually, the exchange interaction between atoms and ions is short-ranged, and then the long-ranged exchange interaction comes into existence, which was effective in oxides and fluorides compounds, termed as super-exchange (SE) model. In oxides and fluorides, when 3d TM ions and intermediates do the electron hybridization by the diamagnetic O 2p anion, it causes the virtual hopping of electrons from one to another to create an unpaired electron in these atoms responsible for magnetization in the system. The origin of SE was mainly for AFM coupling because of the favourable kinetic energy advantage, as their interaction through the O 2p orbital stabilizes AFM order, as shown in Fig. 12 (a). Sometimes, SE can be FM, as there is coupling between the half-filled orbital mediated by oxygen to empty orbital and, because of Hund's rule, polarization favours the energy advantage leading to FM order [29], as shown in Fig. 12 (b). Also, SE depends upon the exchange energy ($\mathscr{J}$) described by Heisenberg Hamiltonian and the bond angle M-O-M and the interatomic distance.

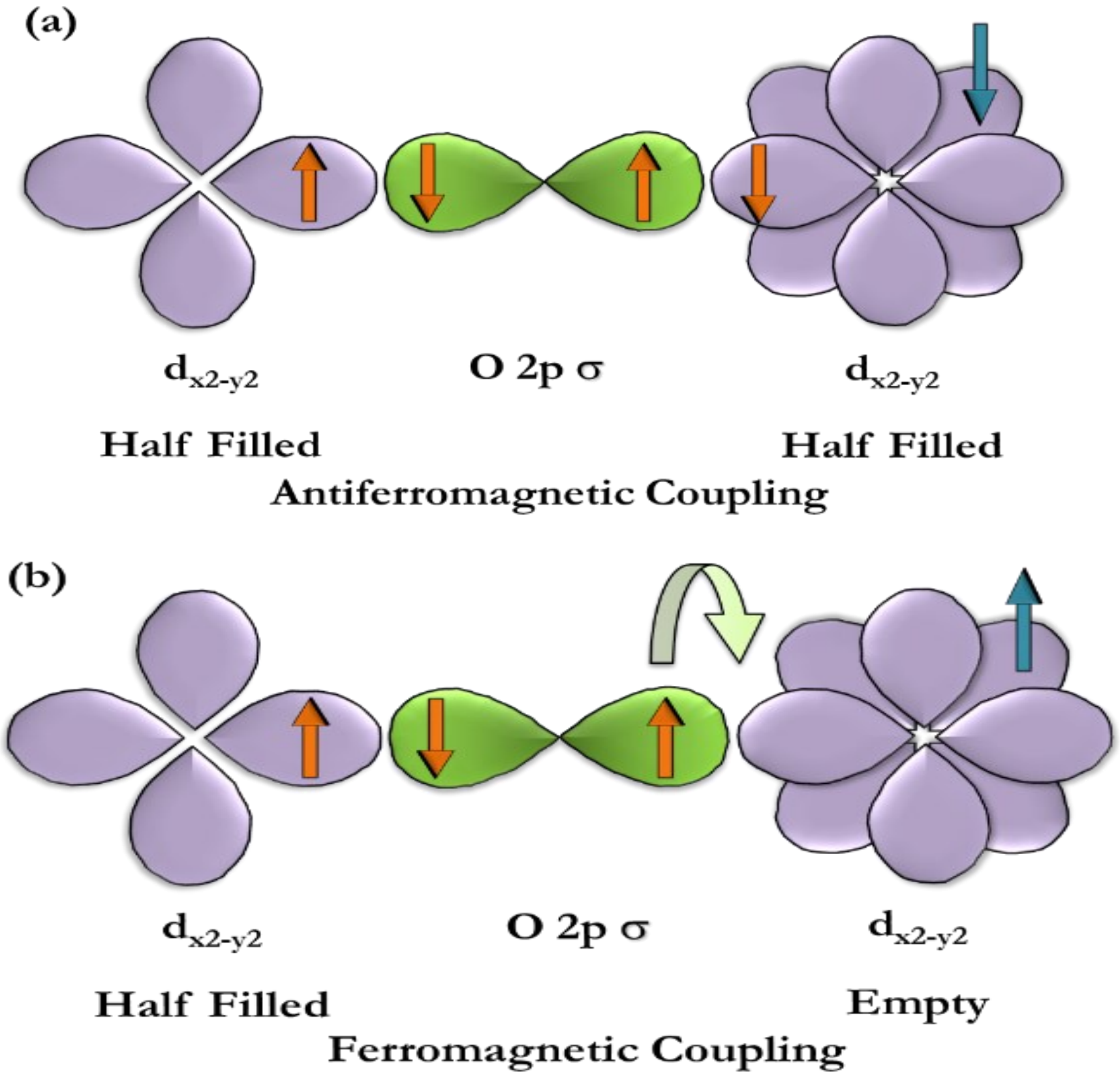

**Figure 12.** Depiction of super exchange interaction in oxides as **(a)** Antiferromagnetic and **(b)** Ferromagnetic coupling.

### 4.5 Bound Magnetic Polarons (BMP) Model

The earlier mechanism to observe FM order depended on the carrier concentration or the exchange interactions, but the low carrier concentration and non-observant of long-range FM order in oxides and insulators were unable to explain the origin of magnetism in DMOs. To overcome this shortcoming, Kaminski et al. proposed a new theoretical model to observe long-range FM order in semiconductors and further improved and explained by Coey et al. (2005) specifically for semiconducting oxides, known as Bound Magnetic Polarons (BMP) model [33,34].

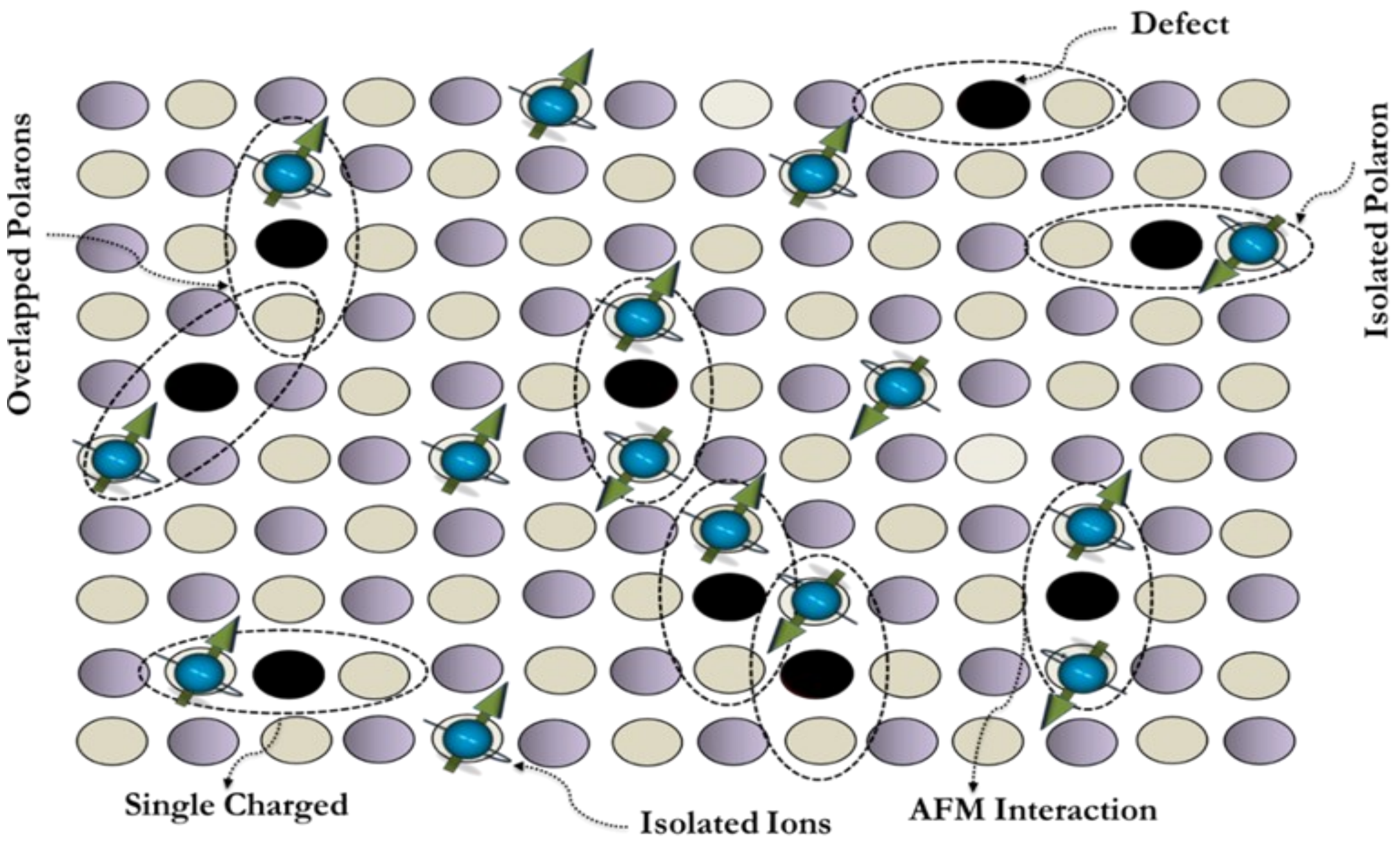

**Figure 13.** Demonstration of BMP and their interaction.

In oxides materials, when sufficient spins localized because of the substitution of TM ions, they interact with weekly bound carriers or defects which were trapped within the limits of effective Bohr radius, behaved as 1s hydrogenic orbital resulting in a polarization of spins, and formed significant magnetic moment by coupling with nearer atoms, leading to the formation of BMP. The basic representation of the BMP model is illustrated in Fig. 13. The BMP model depends on the change of temperature. With the decrease in temperature, the size of the polaron increases. When the size of the BMP becomes large enough, the long-range FM order is achieved. The more significant number of defects, carriers, and high doping lead to forming of a greater occupied volume of BMP, which overlaps with adjacent BMP and increases FM order in the system. The BMP model is integrally attractive to the low carrier density system, which makes the BMP model attractive to both *p*-type and *n*-type materials [35]. Also, BMP theory suggests the radius of polarons is small, and if the dopant level crosses the fermi level, then high $T_c$ can be achieved.

**4.6 $d^0$ magnetization**

The ambiguity on the origin and mechanism of DMOs came to a point when Coey et al., in 2004, observed magnetization in undoped $HfO_2$ despite its diamagnetic characteristics, which opened the path to fabricate new oxide-based semiconductor without the injection of magnetic impurity, termed as "$d^0$ magnetization" [34]. After that, many oxides, such as undoped ZnO, $TiO_2$, $In_2O_3$, $Al_2O_3$ etc., observed the $d^0$ magnetization phenomenon. The observation of magnetism in these nonmagnetic oxides is possibly due to the observance of lattice or surface defects since the above-stated materials are nonmagnetic in bulk when they are defect free. The observation of surface or lattice defects mostly depends upon the valence state of cation and anions, growth conditions and sintering temperatures, gas flow etc. the controversiality of $d^0$ magnetization in oxides leads to the use of highly sensitive characterization techniques such as X-ray-absorption near-edge structure (XANES) spectroscopy, X-ray magnetic circular dichroism (XMCD) [36]. These techniques are essential to eliminate the possibility of secondary phases or to detect defect states, cation, or anion vacancies.

## 5. DMOs based Spintronics Devices

The success of GMR/TMR opened the paths for new potential spintronics applications, which prompted the development of semiconducting spintronics devices. As the dominance of the semiconductor market makes it more attractive and combined with spintronics applications. DMO-based devices are listed below:

### (a) Spin FET

Datta and Das et al. demonstrated a spin-based field-effect transistor on GaAs semiconductors. They expected that the spin-polarized electron emitting from the source would pass through the narrow channel and collect through the spin drain. The configuration of the Spin FET device is like today's transistors, which employ spin injection and detection as properties of source and drain and can be influenced by the gate voltage. The gate voltage changes the angle of spin precision, which can induce the alignment of spins in semiconducting channels. Thus, manipulating gate voltage resulted in an on-off state depending upon spin alignment [37]. Later, Sato et al. proposed spin FET for DMOs material, as shown in Fig.14, in which gate modulation and spin injection are based on the injection of holes [38][39].

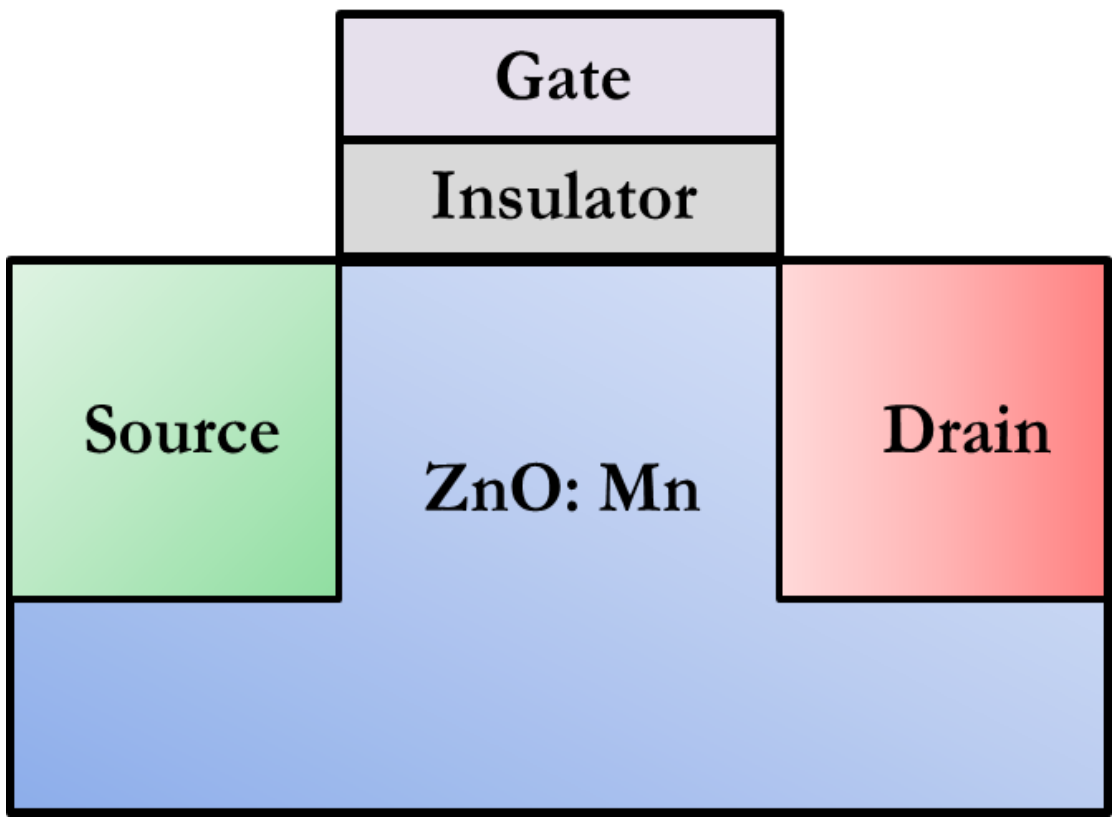

**Figure 14.** Schematic representation of Spin based FET based on DMOs

Spin FET is expected to occur at lower voltages, resulting in low power consumption than conventional FET, which can lead applications to very low power consumption microprocessors with more productivity.

### (b) Spin LED

Ohno et al. fabricated spin-light-emitting diodes of Mn-substituted GaAs semiconductors on GaAs substrate using molecular beam epitaxy (MBE). The spin polarization was done by injecting holes in DMS material and unpolarized electrons from the non-magnetic side, which recombine in quantum wells. The spin polarization of holes, measured by the intensity of left and right circularly polarized light ($\sigma^+$) of emitted electroluminescence (EL) separated by the distance d. The EL studies reveal hole spin polarization in the magnetic semiconductors greater than 200 nm, having a coercivity of 40 Oe between the p-n junction.

Fig. 15 shows the model structure of GaAs-based spin LED. The observation of Spin-LED below 52 K acknowledged the electric injection of spins in a semiconductor without applying magnetic fields, which can eliminate the need for a polarizing filter in conventional devices. [40].

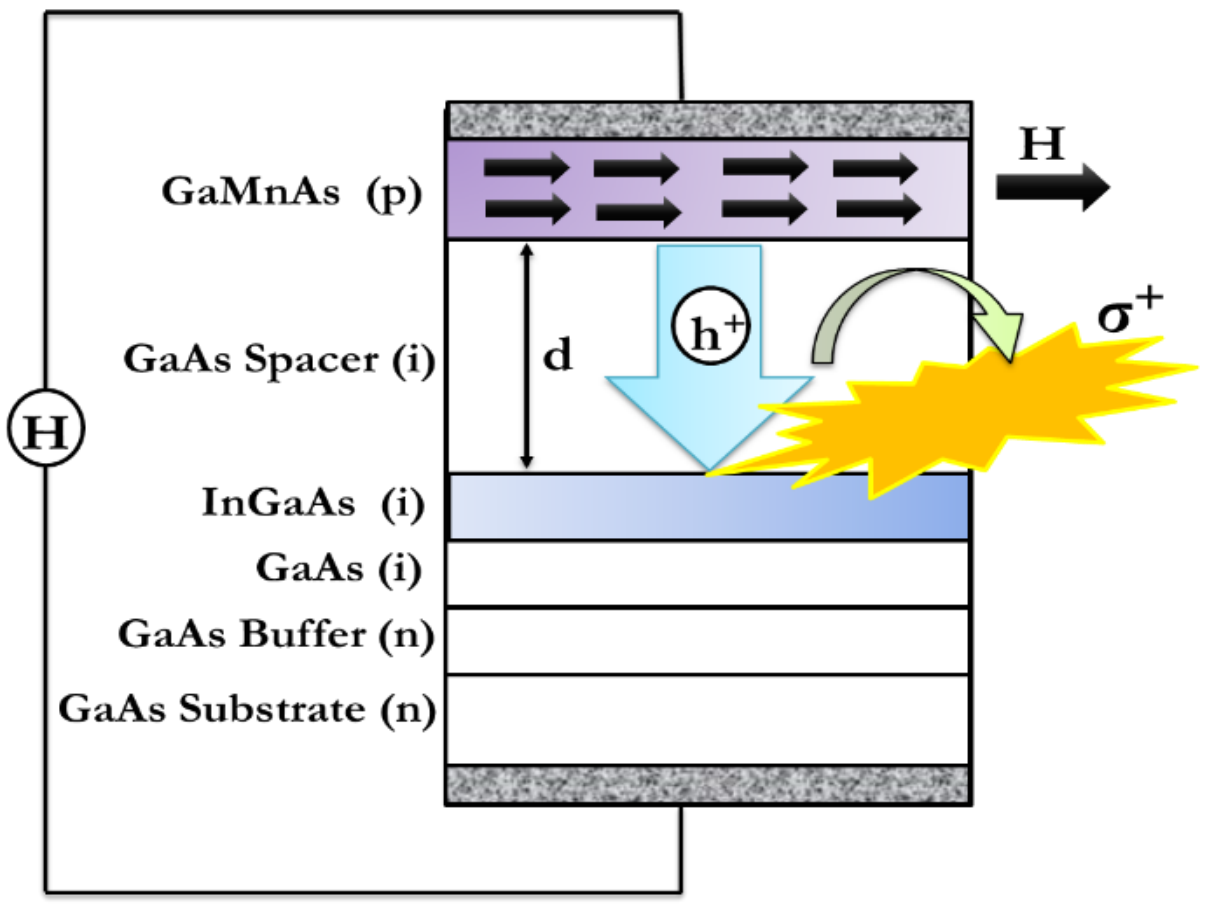

**Figure 15.** DMS based spin LED device structure fabricated by Ohno and group.

### (c) Transparent Ferromagnetic

Sato and Yoshida et al. proposed the photo-based magnetic structure of TM ions doped ZnO to make a half-metallic ferromagnetic state with hole/electron doping fabricated in GaAs substrate, as shown in Fig. 16. They hypothesized that introducing TM ions based on the oxidation states results in electron or hole doping in ZnO semiconductors. These electrons or holes can be influenced by the photoexcitation on the interface of the ZnO and GaAs band gap, which can result in the favourable energy of photons between them, detected by the magneto-optical effect. The use of semiconducting oxides such as ZnO or $Cr_2O_3$ having transparency at visible light and doped magnetic ions can result in spintronics devices and open the path of a transparent DMS device in which we can control spin, electron, and photon [39].

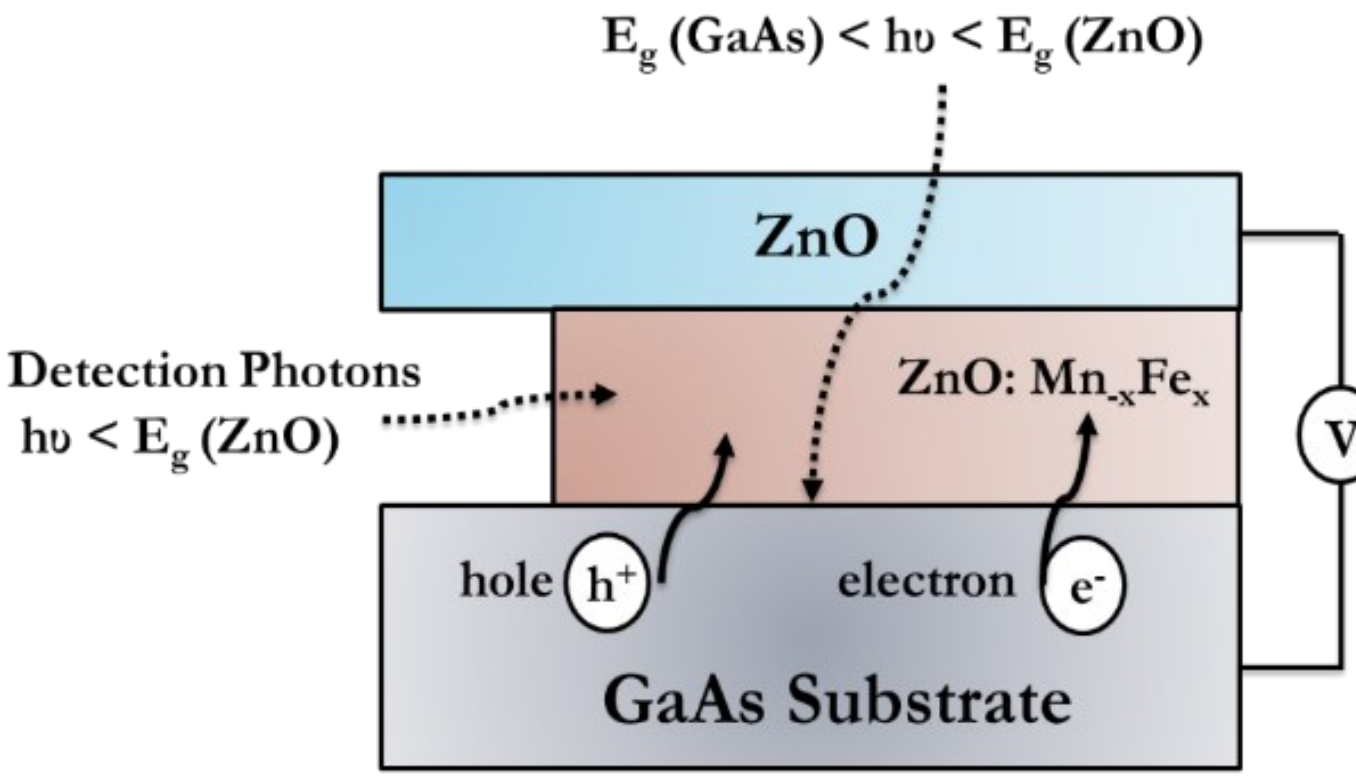

**Figure 16.** Schematic of transparent photomagnetic.

### (d) Spin Solar Cell and Photodiode

Endres et al. fabricated a spin solar cell and observed the spin photodiode effect on (Ga,Mn)As on a GaAs substrate using the MBE technique. They introduced the concept of spin extraction and injection, which combines the principles of a solar cell with spin accumulation (a buildup of electrons with aligned spins in a specific region). They demonstrated that illumination with unpolarized light on a *p*-type ferromagnetic (FM) p-n junction can achieve optical spin injection. The working principle of the spin solar cell is based on the induction of light-based photovoltage, which drives a tunnelling current through a gap in the material. Photoexcited electrons in this process are weakly spin polarized. Spin extraction occurs as these weakly spin-polarized electrons tunnel, resulting in a buildup of spin on one side of the cell. The spin photodiode effect utilizes a reverse bias to control tunnelling and generate spin-polarized photoexcited electrons that accumulate with an opposite orientation compared to spin extraction. [41]. This opens the path of devices not limited to converting incident light into voltage but also for spin accumulation without using the direct band gap of traditional semiconductors, as shown in Fig. 17.

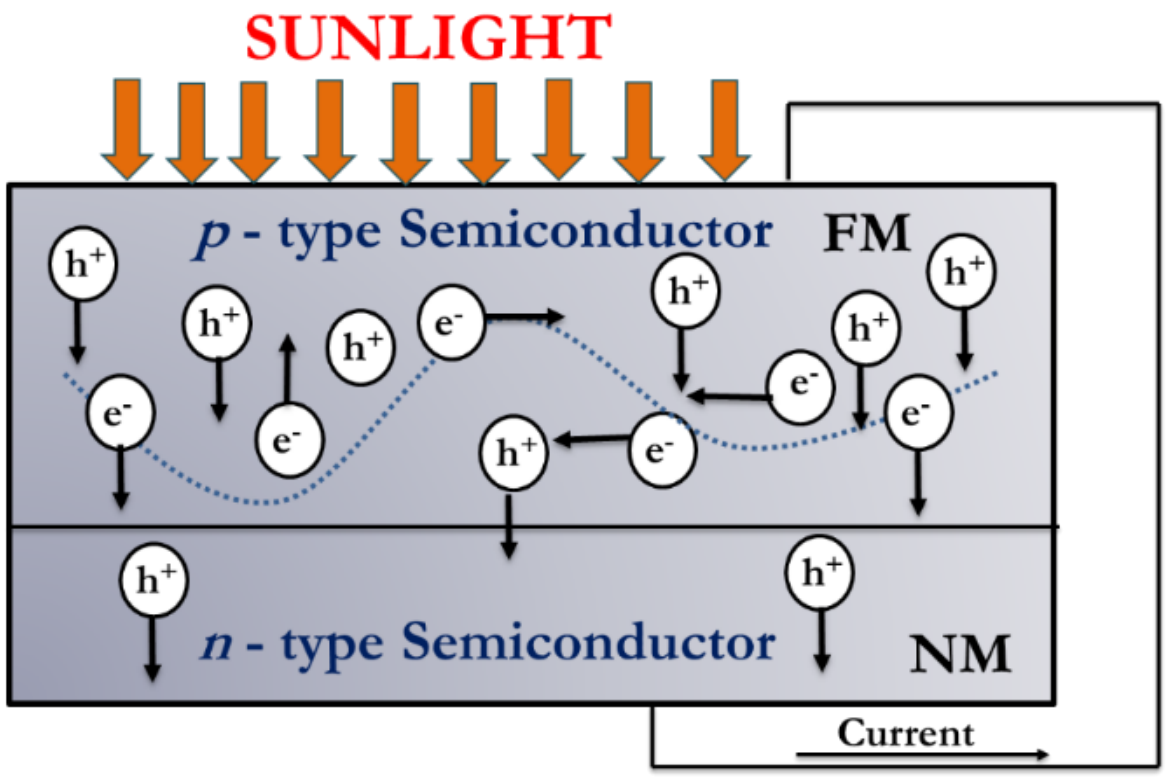

**Figure 17.** DMS based spin solar cell and photodiode.

## 6. DMOs Based Semiconducting Oxides

Dietl et al. created a theoretical model that a small concentration of Mn (x = 0.05) doping in ZnO can achieve *p*-type conductivity with a wide optical band gap and high $T_c$ due to hole-mediated exchange interactions [21]. In 2001, Matsumoto et al. deposited 7% Co-doped $TiO_2$ thin films using MBE and achieved a transparent FM semiconductor with appreciable n-type conductivity [23]. The first oxides which obtained high $T_c$ were thin films later, afterwards DMOs phenomenon also observed in nanoparticles and nanostructures. Lately, attention was focused on the oxide materials due to their unique natural properties, such as excellent optical, chemical, and physical properties. As discussed in the previous section regarding selecting suitable DMOs materials, they obey all mentioned criteria. The temptation to achieve semiconducting RTFM in oxides with material-specific properties makes the status of DMOs materials so intriguing among researchers worldwide. Several attempts have been made to achieve the status of DMOs to obey all the requirements of DMOs for future semiconductor spintronics devices.

An extensive literature review has been done based on *p*-type and *n*-type oxides using research databases such as Web of Science and Scopus by adding specific keywords based on magnetic semiconductors from the inception of DMSs from 1990 to 2024. Fig. 18 shows the trends of published research articles in DMS and DMOs.

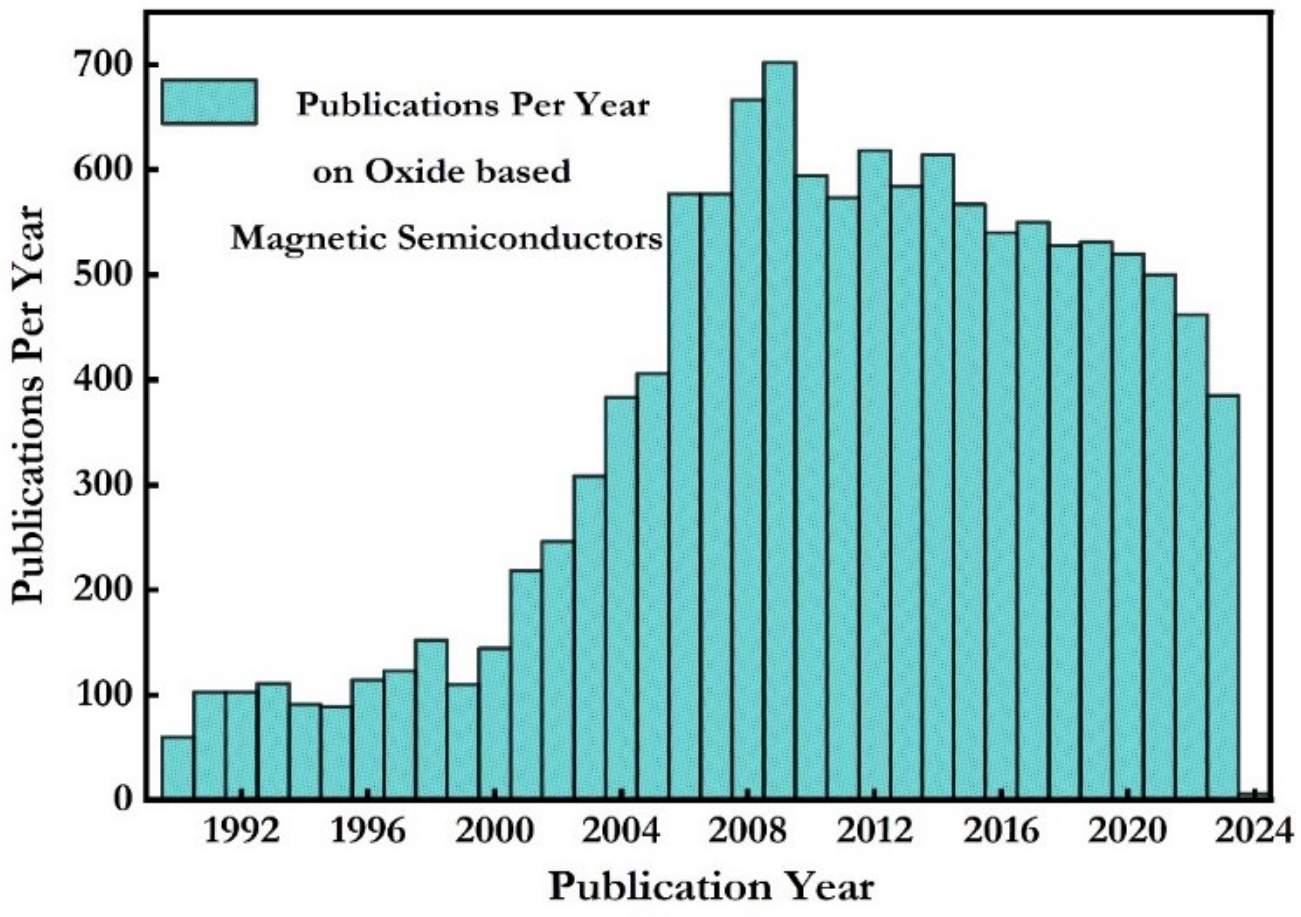

**Figure 18.** Number of publications per years on DMOs, according to exported web of science database.

At present, there are over 12,000 publications that reported RTFM in oxides, and a maximum of them belong to *n*-type oxides despite being theoretically confirmed in p-type oxides first. Tables 1 and 2 summarize the details of various DMO systems in terms of synthesis and fabrication as nanoparticles and thin films, respectively, with TM ions threshold, magnetism value, transition temperature with originated conductivity, and optical band gap with magnetic origin.

In contrast, Fig. 19. shows further bibliographic analysis regarding the density of keywords related to magnetic semiconductors. For bibliographic analysis, the 1232 research articles' abstracts and titles were exported in text form from the Web of Science databases.

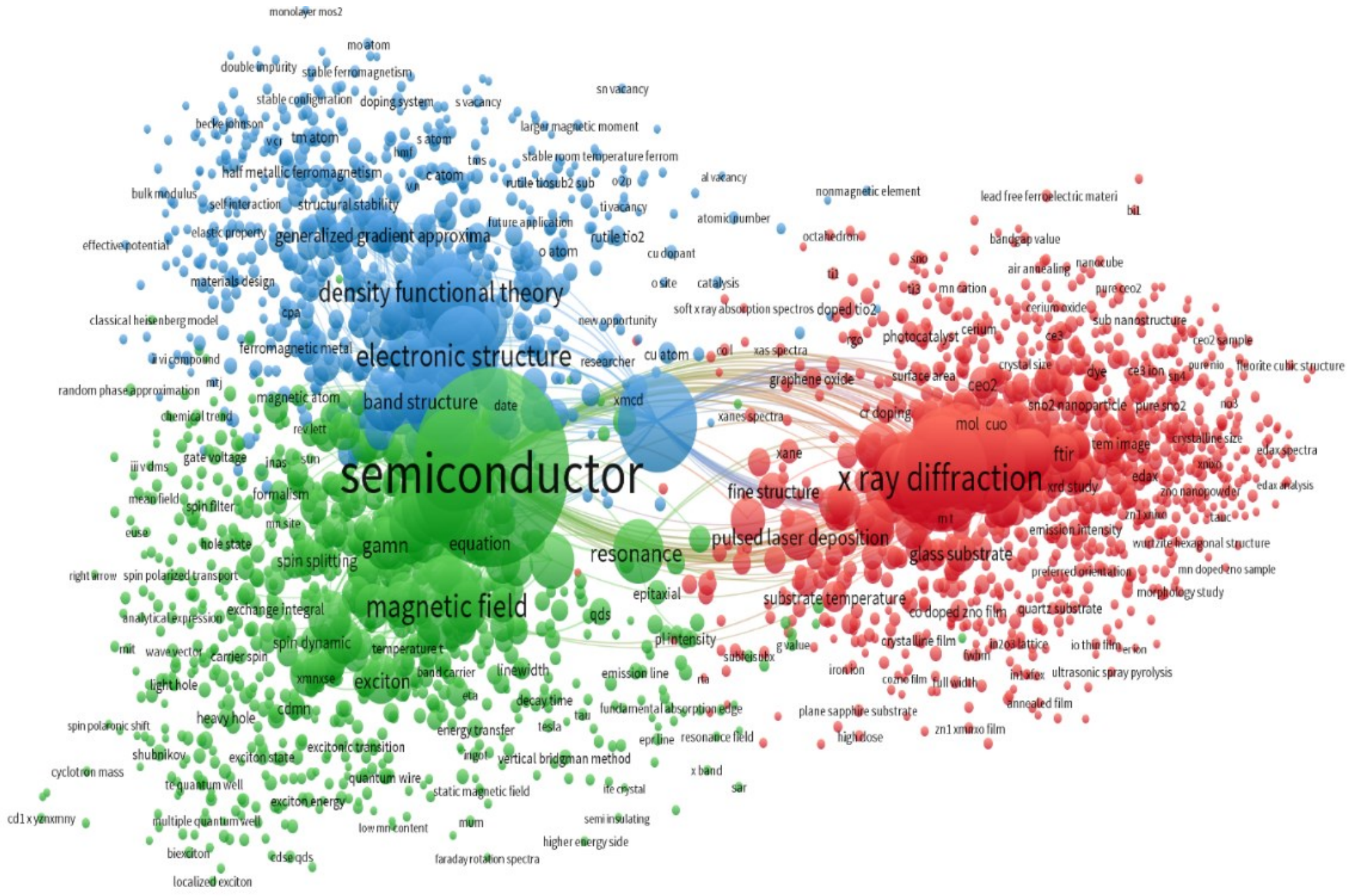

**Figure 19.** Important keywords associated with oxide based magnetic semiconductors analysed using VOS software.

**Table 1.** The details of some DMO nanoparticles materials, with their synthesis techniques, TM ions threshold, magnetism value, transition temperature, and optical band gap with magnetic origin.

| Material | Synthesis Method | TM ions | Doping Threshold | Magnetic Moment per ion at RT | $T_c$ / $T_n$ (K) | Optical Bandgap (eV) | Origin of Magnetism | Ref |
|---|---|---|---|---|---|---|---|---|
| **ZnO** | Sol-gel | $Co^{2+}$<br>$Mn^{2+}$ | 0.05 | 0.35 µb /Co<br>- | RTFM | - | *n*-type conductivity stabilizes FM | [42] |
| **$In_2O_3$** | Chemical vapor transport | $Cr^{3+}$ | 0.03 | 0.074 µb /Cr | ~ 375 K | - | Oxygen vacancies leads to formation of BMP | [43] |
| **$TiO_2$** | Sol-gel | $N_2$ | 0.06 | ~ $5.8 \times 10^{-6}$ µb /N | RTFM | _ | Oxygen vacancies | [44] |
| **$SnO_2$** | Wet chemical | $Fe^{3+}$<br>$Ni^{2+}$ | 0.05 | 0.014 µb/Fe<br>0.013 µb/Ni | RTFM | 3.87<br>3.88 | FM super-exchange interactions | [45] |
| **$CeO_2$** | Sol-gel | $Fe^{3+}$ | 0.01 | 0.003 µb /Fe | 300 K | 3.08 | Intrinsic defects exchange interactions | [46] |
| **ZnO** | Co-precipitation | $Mn^{2+}$ | 0.05 | $7.2 \times 10^{-3}$ µb /Mn | RTFM | - | p-d hybridization of Mn ions | [47] |
| **$TiO_2$** | Sol-gel | $Cu^{2+}$ | 0.04 | 0.0024 μb/Cu | < 320 K | _ | Oxygen vacancies and Cu exchange interactions | [48] |
| **CuO** | Sol-gel | $Fe^{3+}$ | 0.08 | 0.27 $\mu_b$ /Fe | < 350K | - | Shape Anisotropy | [49] |
| **$CeO_2$** | Co-precipitation | $La^{3+}$ | 0.01 | 1.834 $\mu_b$ /La | 380 K | 3.5 - 6 | Giant orbital Para magnetism | [50] |
| **$SnO_2$** | Co-precipitation | $Cr^{3+}$ | 0.01 | $6.2 \times 10^{-4}$ μb /Cr | RTFM | 4.26 | FM super exchange interactions | [51] |
| **$TiO_2$** | Solid state | $Mn^{2+}$ | 0.02 | $4.32\times 10^{-4}$ μb /Mn | < 420K | - | Oxygen defects and segregated Mn | [52] |
| **$HfO_2$** | Metathesis | $Y^{3+}$ | 0.10 | $1.08\times 10^{-4}$ μb /Y | RTFM | - | *n*-type doping and creation of oxygen vacancies | [53] |
| **CdO** | Co-precipitation | $Mn^{2+}$ | 0.05 | 0.0056 μb /Mn | RTFM | 3.87 | FM exchange coupling | [54] |
| **$Cr_2O_3$** | Wet-chemical method | $Ga^{3+}$ | 0.045 | 2.10 $\mu_b$ /Ga | 430 K | 2.85 | Transport of charge carrier and Frustrated surface spins | [55] |
| **$SnO_2$** | Combustion Method | $Cr^{3+}$ | 0.02 | 6.6× $10^{-3}$ µb /Cr | < 300 K | 3.9 | Oxygen vacancies mediated BMP model | [35] |
| **NiO** | Co-precipitation method | $Ce^{3+}$ | 0.05 | 0.018 µb /Ce | RTFM | 3.55 | Oxygen vacancies mediated BMP | [56] |
| **$In_2O_3$** | Solid-state route | $Co^{2+}$ | 0.07 | 1.5 µb /Co | 718 K | _ | Occurrence of $Co^{2+}$ ions and BMP model | [57] |
| **ZnO** | Solid state | $Ni^{2+}$<br>$Cu^{2+}$<br>$Fe^{3+}$<br>$Mn^{2+}$ | 0.05 | 0.0034 µb /Ni<br>0.00029 µb /Cu<br>0.0128 µb /Fe<br>0.0023 µb /Mn | < 300 K | 2.95<br>3.15<br>3.55<br>3.70 | Oxygen Vacancies, charge carriers and mixed-valence states | [58] |
| **$Al_2O_3$** | Precipitation method | $Ni^{2+}$ | 0.04 | 5.54× $10^{-7}$ µb /Ni | RTFM | 3.72 | Ni vacancies at dislocation core of $Al_2O_3$ | [59] |
| **$CeO_2$** | Co-precipitation method | $Ni^{2+}$<br>$Co^{2+}$ | 0.05 | 0.138 µb /Ni<br>0.244 µb /Co | ~ 320 K<br>~ 330K | 2.0<br>1.6 | Oxygen vacancies induced FM | [60] |
| **$TiO_2$** | Co-precipitation method | Fe: Ga – H<br>Fe: Ni – H | 0.01 - 0.05 | -<br>$4.13\times 10^{-6}$ μb /Fe | PM<br>RTFM | 3.3<br>3.1 | Oxygen vacancies and super exchange interactions | [61] |
| **$Cr_2O_3$** | Co-precipitation method | $Ni^{2+}$<br>$Fe^{3+}$<br>$Mn^{2+}$<br>$Co^{2+}$ | 0.07 (0 to 0.07)<br><br>0.05 (0 to 0.10) | 1.12 $\mu_b$ /Ni<br>4.52 $\mu_b$ /Fe<br>-<br>- | RTFM<br>240 K<br>300 K<br>301 K | 2.62<br>2.56<br>2.70<br>2.85 | BMP<br>Oxygen defects<br>Spin frustrations.<br>Spin canting | [62–64]<br>Our Work |

**Table 2** The details of some DMO materials, with their fabrication methods, TM ions threshold, magnetism value, transition temperature with originated conductivity, and optical band gap with magnetic origin.

| Material | Fabrication Method | TM ions | Doping Threshold | Magnetic Moment per ion at RT | $T_c$ / $T_n$ (K) | Optical Bandgap (eV) | Carrier Types | Origin of Magnetism | Ref |
|---|---|---|---|---|---|---|---|---|---|
| **$TiO_2$** | MBE | $Co^{2+}$ | 0.07 | 0.32 μb /Co | < 400 K | 3.1 | *n-type* | Mixing of Co granular precipitates | [23] |
| **ZnO** | PLD | $Co^{2+}$ | 0.2 | 2 μb /Co | 280 K | _ | *n-type* | RKKY interactions | [65] |
| **$In_2O_3$** | Reactive Evaporation | $Cr^{3+}$ | 0.02 | 1.5 μb /Cr | 930 K | 3.75 - 3.95 | *n-type* | Magnetic polarons | [66] |
| **ZnO** | Direct chemical synthesis | $Co^{2+}$ <br> $Mn^{2+}$ ($N_2$) | 0.035 <br> 0.02 | 0.04 µb/Co <br> 1.2 µb /Mn | < 300 K <br> < 300K | _ <br> _ | *n-type* <br> *p-type* | Dopant p-d conductive hybridization | [67] |
| **$In_2O_3$** | PLD | $Co^{2+}$ | 0.01 | 0.5 µb /Co | 400 K | _ | _ | Oxygen vacancies | [68] |
| **$SnO_2$** | Spray pyrolysis | $Mn^{2+}$ | 0.075 | 0.18 µb /Mn | 550 K | 3.8 | *n-type* | Defects mediated exchange interactions | [69] |
| **$In_2O_3$** | PLD | $V^{2+}$ | 0.05 | 1.7 µb/V | < 300 K | _ | _ | Effect of $O_2$ partial pressure | [70] |
| **$TiO_2$** | PLD | $Fe^{2+}$ | 0.01 | 6.9 µb/Fe | RTFM | _ | _ | Charge transfer | [71] |
| **$CeO_2$** | PLD | $Cu^{2+}$ | 0.03 | 1.1 µb /Cu | RTFM | 3.43 | _ | Mixed valence states of Cu | [72] |
| **$ZrO_2$** | PLD | $Mn^{2+}$ | 0.05 | 13.8 µb /Mn | < 400 K | _ | _ | Defects mediated exchange interactions | [73] |
| **$SnO_2$** | RF Sputtering | $Li^{+}$ | 0.12 | 0.028 µb /Li | ~ 300 K | 3.9 | *n to p-type* | p – p coupled Exchange interactions ($d^0$ FM) | [74] |
| **CuO** | RF Sputtering | $Mn^{2+}$ | 0.298 | 0.15 µb /Mn | 99 K | _ | *p-type* | Mn and Cu ions interactions and cations vacancies | [75] |
| **$TiO_2$** | DC Sputtering | $Mn^{+}$ | 0.05 | 0.80 µb /Mn (20 K) | 230 K | _ | *n-type* | High carrier concentrations and originated defects | [76] |
| **$In_2O_3$** | PLD | $Fe^{3+}$ | 0.18 | 0.125 µb /Fe | < 300 K | _ | *n-type* | Spin-polarized charge carriers | [77] |
| **$SnO_2$** | PLD | $Ga^{3+}$ | 0.02 | 0.025 µb /Ga | ~ 350 K | 4.18 | *n-type* | Carrier mediated interactions | [78] |
| **MgO** | Sputtering | -- (STO substrate) | _ | 0.056 µb /Mg | < 300 K | _ | *n-type* | Mg vacancies and carrier interactions ($d^0$ FM) | [79] |
| **$CeO_2$** | RF Sputtering | $Mn^{2+}$ | 0.15 | $2.2 \times 10^{-6}$ µb /Mn | RTFM | _ | _ | Oxygen vacancies mediated interactions | [80] |
| **NiO** | Electro-deposition | Cu | 0.03 | $4.55 \times 10^{-7}$ µb /Cu | RTFM | 3.20 | *Dielectric* | Particle size and domain wall motion | [81] |
| **CdO** | Perfumed atomizer | $Ba^{2+}$ | 0.03 | $6.02 \times 10^{-7}$ µb /Ba | RTFM | 2.5 | _ | Oxygen vacancies mediated FM | [82] |
| **$Cr_2O_3$** | PLD | $Ti^{4+}$ | 0.02 | 0.068 $\mu_b$/Ti | RTFM | _ | _ | Oxygen vacancies | [83] |
| **ZnO** | MBE | $As^{+}$ ions implantation | -- | 0.035 $\mu_b$/Zn | RTFM | 3.2 | *Resistive* | Lattice induced defects | [84] |
| **NiO** | Sol-gel spin coating | $Mn^{2+}$ | 0.15 | 0.39 $\mu_b$/Mn | $<$ 300 K | 3.57 | _ | Oxygen vacancies induced BMP | [85] |
| **ZnO** | RF Sputtering | $Cu^{2+}$ | 0.05 | 1.87 $\mu_b$/Cu | RTFM | -- | *n-type* | Mixed valence state formation of BMP | [86] |
| **$TiO_2$** | DC Sputtering | $Co^{2+}$ | 0.12 | $1.5 \times 10^{-6}$ μb /Co | 350 K | ~ 3.57 | _ | Dipolar Co grains interactions | [87] |
| **$Cr_2O_3$** | PLD | $Ni^{2+}$<br>$Mn^{2+}$<br>$Co^{2+}$ | 0.05 | 0.02 $\mu_b$/Ni<br>0.05 $\mu_b$/Mn<br>0.024 $\mu_b$/Co | 343 K<br>324 K<br>350 K | 3.57<br>3.58<br>3.56 | *p-type* | Charge transfer and Indirect RKKY interaction | [88]<br>OurWork |

The VOS software analysed the exported database views to determine the relationship between the various interlinked keywords. The density of colour and the size of the text reveal the relation between the keywords. The database illustrates the precise relationship between the semiconductor (materials) and x-ray diffraction (characterization). As we discussed earlier, the control and origin of DMS/DMO still need to be clarified, making it essential to use various techniques to find the origin of DMO. The comprehensive different oxides literature survey based on the conductivity are discussed below:

## 6.1 *n*-type based DMOs.

### 6.1.1 Titanium Oxide ($TiO_2$)

In 2001, Matsumoto and the group observed the first oxide based DMS with the deposition of Co-doped anatase $TiO_2$ thin films using MBE with 7 % solubility. They obtained a transparent RTFM semiconductor with appreciable *n*-type conductivity. The observation of RTFM is attributed to the mixing of Co granular nanoclusters [23]. Years later, in 2003, S. R. Shinde and colleagues examined the optical, transport, magneto-transport and magnetic properties of Co-doped anatase $TiO_2$ thin films deposited by PLD techniques. They observed $T_c$ of 1123 K (x = 0.03), which reduced to 700 K (x =0.07) because of Co clusters having a crystallized size of 20-50 nm. Also, all the thin films show high transparency, with n-type conductivity, due to oxygen vacancies [89].

In 2005, S. Duhalde and colleagues explored non-magnetic Cu-substituted $TiO_2$ thin films of 10 at% concentration grown on LAO substrate by the PLD technique to further understand the origin of magnetism. Surprisingly, they observed RTFM with a higher magnetic moment of 1.5 $\mu_b$/Cu, attributed that oxygen vacancies play a crucial role in the observation of RTFM [90]. Similarly, D. Lu et al. (2008) grew epitaxial Cu (10, 15, 20 %) and substituted $TiO_2$ thin films on $SiO_2$ substrates using reactive magnetron sputtering. They also observed RTFM, with a $T_c$ of 350 K, and concluded that the observation of magnetization is due to the oxygen vacancies and their interactions between the neighbouring Cu atoms [91].

Coey and group (2010) grew Fe-doped rutile $TiO_2$ thin film up to 5 % concentration on an r-cut sapphire substrate using the PLD method. They observed a wide optical band gap of 3.9 - 4.2 eV due to the formation of the Magnéli phase and a very high magnetic moment of 6.9 $\mu_b$/Fe atom (1%). The observation of this magnetization was neither from Fe nor Fe-based secondary phase. Mössbauer spectra were done to determine the role of the mixed valence state of Fe and Ti, resulting in the formation of a new model to explain magnetism known as Charge transfer ferromagnetism (CTF). CTF is a defect-related FM model, where the spin split defect band is populated by the charge transfer from the charge reservoir mixed oxidation state of ions in a crystal lattice, as shown in Fig. 20 [71].

Y. J. Lee and group (2010) investigated the effects of spin tunnelling by employing 1.4% Co doped $TiO_2$ as an interfacial layer in epitaxial LSMO/STO heterostructures as magnetic tunnel junctions (MTJ) grown using PLD on STO substrate. They studied the effect of an interfacial layer on tunnel magnetoresistance (TMR) studies and their bias dependence and observed a reduction in negative TMR with insertion. The decrease in negative TMR was attributed to carrier-induced metallic impurity, which induces RTFM [92].

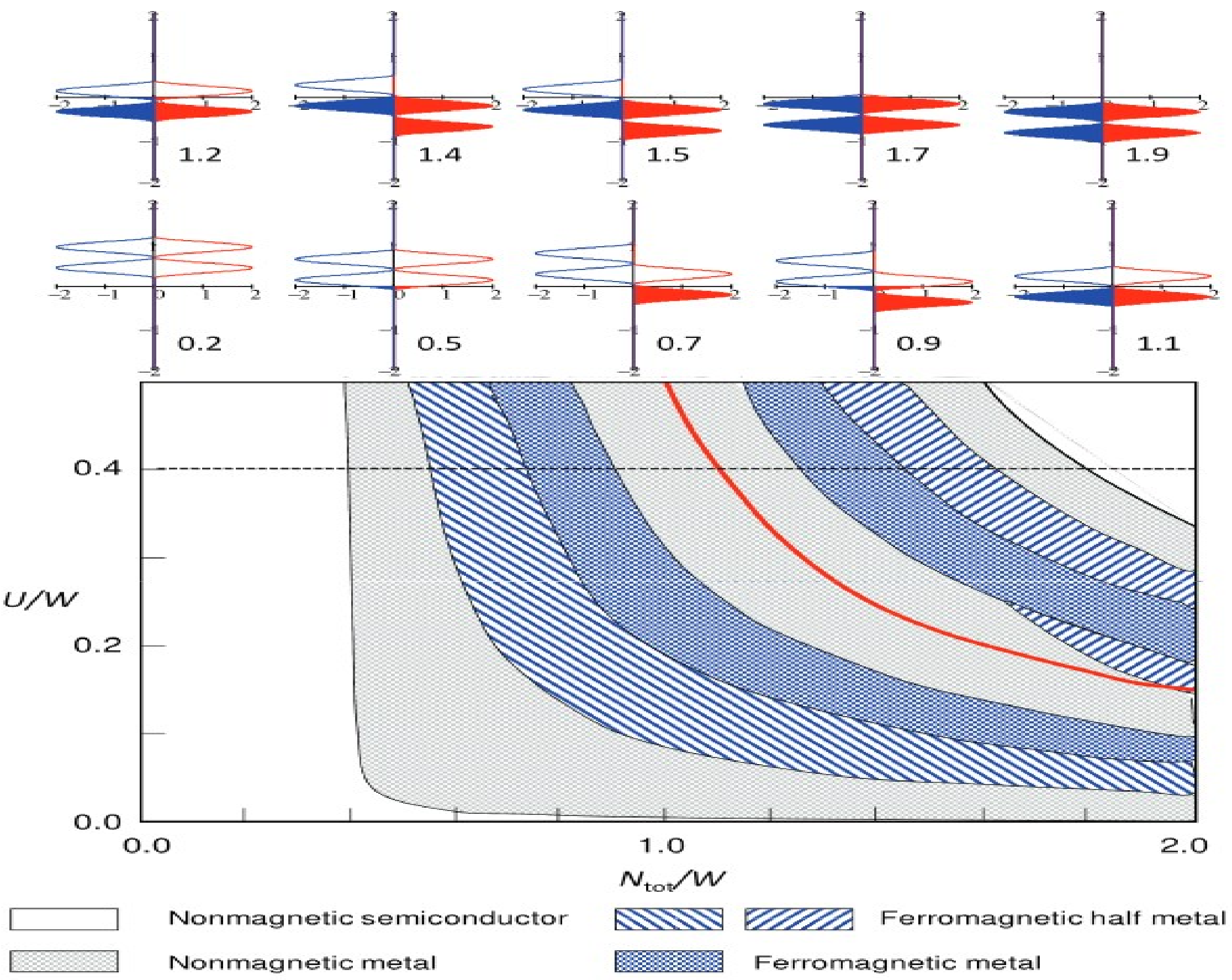

**Figure 20.** Phase diagram theoretically calculated by Coey et al. for magnetic, non-magnetic semiconductors and ferromagnetic metal [71].

H. Thakur and colleagues (2011) grew $TiO_2$ thin films using RF magnetron sputtering. They irradiated heavy $Ag^{15+}$ ions having the energy of 200 MeV with the fluence of $5 \times 10^{12}$ ions/cm$^2$ and observed $d^0$ magnetization confirmed by XAS and XMCD at room temperature. The observation of $d^0$ magnetization is attributed to the unquenched orbital magnetic moment of O 2p shell coupled with nearer ferromagnetic Ti ions due to structural disorder with ions irradiation [36]. A. Rusydi and colleagues (2012) grew technology-critical element Tantalum (Ta) of 5 % concentration on anatase $TiO_2$ using the PLD method with varying sintering temperatures. They reported a transparent thin film, which demonstrates RTFM with n-type conductivity. They also did RBS, XAS, SXMCD and SQUID measurements to validate the results. They concluded that the observed magnetism arises because of the indirect RKKY exchange interaction with free electrons, which occurs due to the cationic vacancies, first verified, and reported in the literature [93].

S. Wang and group (2015) studied the role of interstitial Ti ($V_{Ti}$) vacancies confirmed by strong EPR signal in undoped $TiO_2$ nanostructure prepared by solvothermal treatment method. They observed that the Effect of $V_{Ti}$ causes the shrinking of lattice parameters, which results in the widens of the optical band gap, change in the conductivity from *n*-type to *p*-type with higher mobility and the observation of remarkable photocatalytic properties and RTFM [94]. L. Tseng and group (2016) synthesized the TM (Co, Mn, Ni, and Fe) ions of 1 % concentrations doped in rutile $TiO_2$ nanorods using the molten salt method. They observed the coexistence of diamagnetic-dominated RTFM. They concluded that oxygen vacancies lead to PM contributions, and Ti vacancy leads to FM contributions in the system [95]. B. Parveen and their group studied the role of Ni in $TiO_2$ thin films deposited on the glass substrate using the dip-coating method. They revealed that at 2.5 wt % concentration of Ni, the thin films show RTFM with no external influence, as shown in Fig. 21. The observation of RTFM attributed to trapped F-Centre vacancy induced BMP, which was mediated by localized electrons [96]. V. Akshay and group (2019) studied the optical and magnetic behaviour of Ni-doped anatase $TiO_2$ nanocrystal prepared by the sol-gel method and observed that the narrowing of optical band gap and very weak RTFM with Ni doping due to the formation of oxygen vacancies result in the occurrence of BMP in a system [97]. S. Ravi (2020) studied the effect of Mo dopant on $TiO_2$ nanoparticles prepared using the wet-chemical method up to 1% concentrations. They observed RTFM with a very high $T_c$ of 400 K attributed to the mixed FM valence state of $Mo^{4+}$ [98].

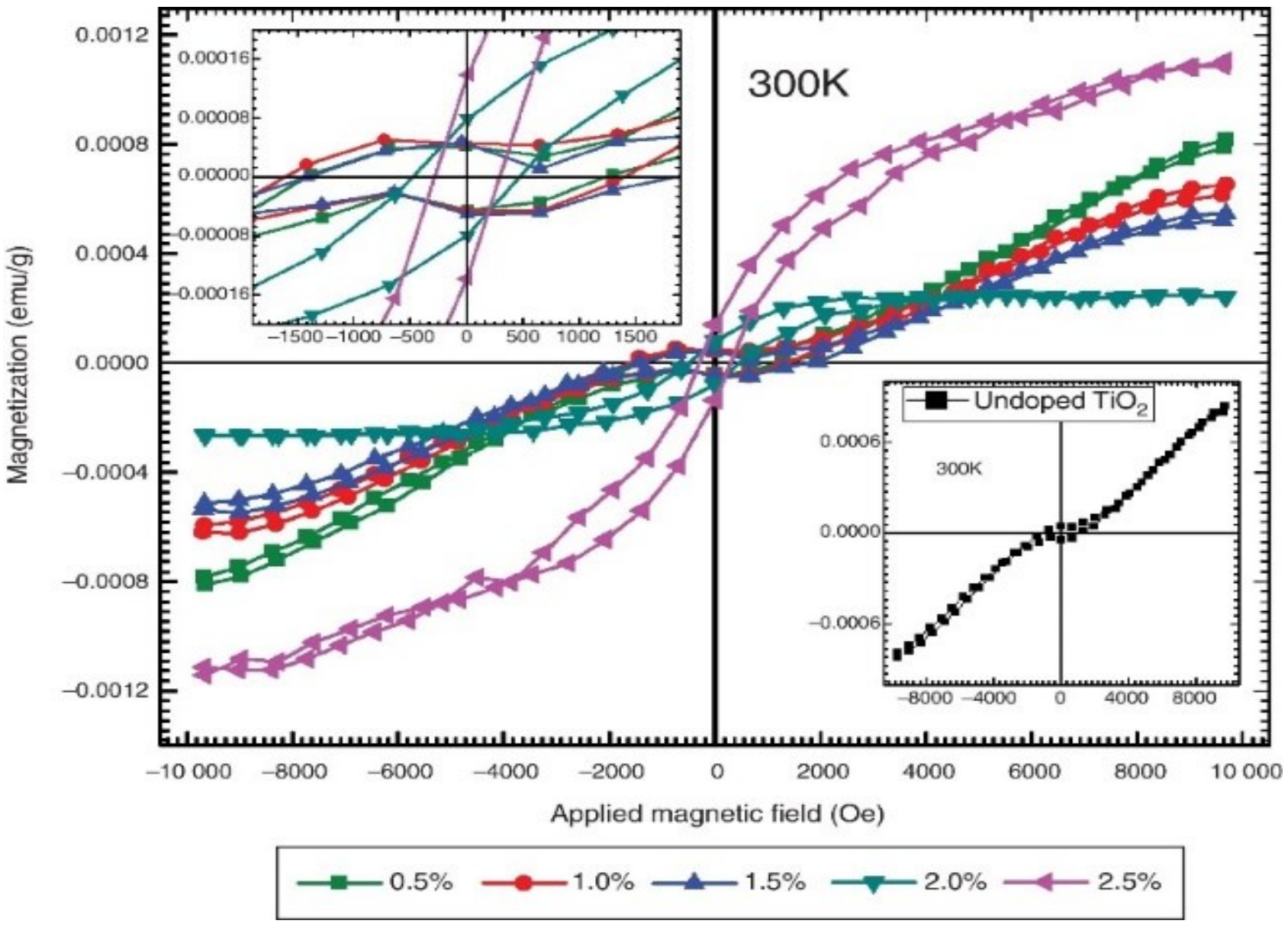

**Figure 21.** The M-H study at 300 K shows RTFM behaviour in all fabricated Ni doped $TiO_2$ thin films with high coercivity attributed to BMP [96].

H. P. Quiroz and colleague fabricate Co/$TiO_2$ and Co: $TiO_2$ bilayer thin films using DC magnetron sputtering on Ti substrate for Magnetic-Resistive Random-access memory(M-RRAM) applications. They employed SET and RESET systems with the effects of the magnetic field, which resulted in low power energy consumption for switching. The I-V study and hall analysis confirm the increase in resistivity up to 234 K, and magneto-resistive (MR) behaviour at room temperature, as shown in Fig.22. The observation of promising characteristics of M-RRAM device is attributed to lattice-free carriers, which were compensated by the oxygen vacancies and Co atoms, as confirmed by the I-V and HRXPS studies [87].

L. Chouhan (2022) et al. doped nonmagnetic Mg up to 0.12 concentration in rutile and anatase $TiO_2$ prepared by solid-state method and observed wide optical band gap and the observation of RTFM with $T_c$ of above 400 K attributed to $d^0$ magnetization occurred due to the oxygen vacancies in $TiO_2$ lattice [99]. Ehsan and group (2023) fabricated a vertical spin-valve device with CoFe/$TiO_2$/CoFe configuration using e-beam lithography. They varied the thickness of interfacial $TiO_2$ to study the suitability of the spin valve with the effect of positive

MR value at low and room temperature and observed a higher TMR of 1.03% at room temperature because of the spin filtering effect due to low $TiO_2$ thickness, which results in $TiO_2$ as favourable oxide for memory storage with low power consumption [100].

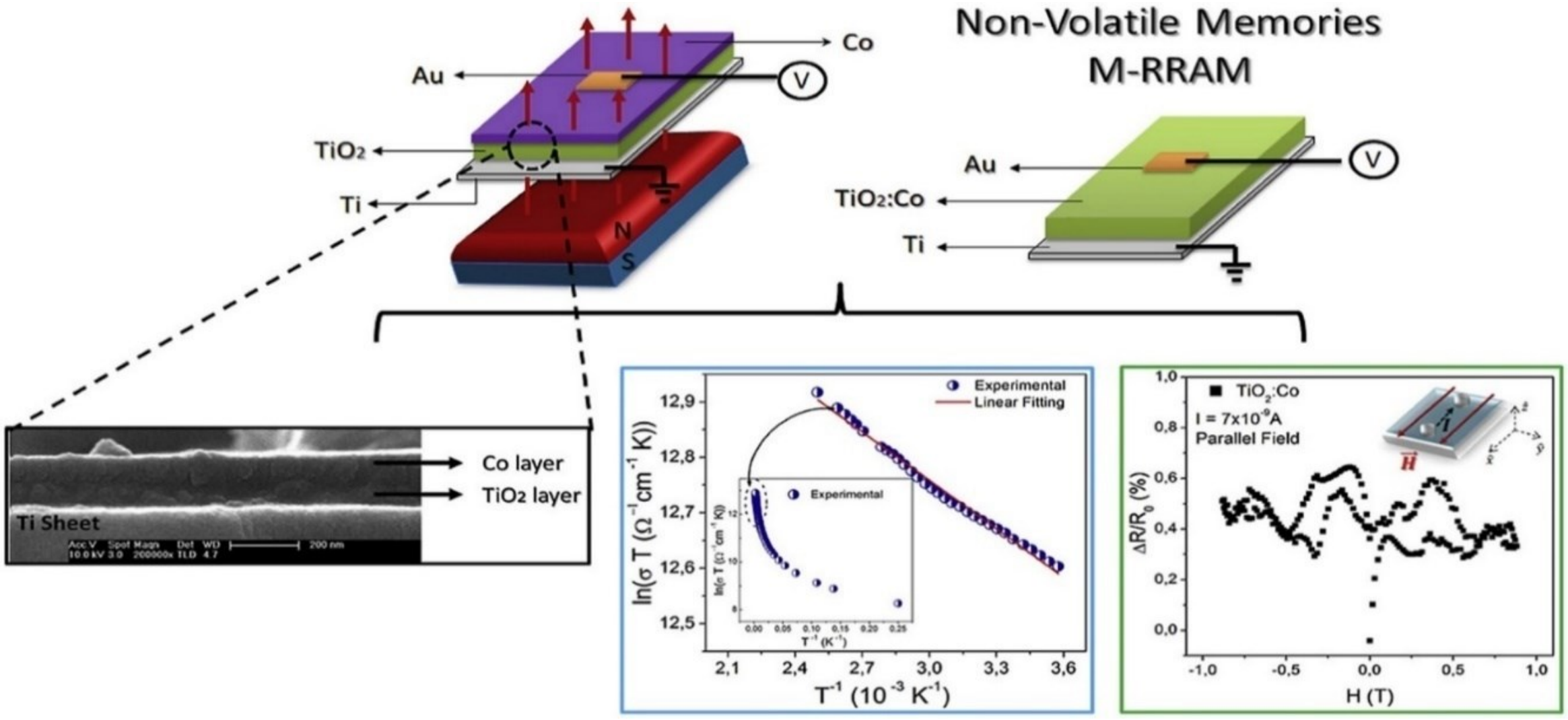

**Figure 22.** The schematics of $Co/TiO_2$ and Co: $TiO_2$ bilayer thin films with or without magnetic field, confirmed by FESEM. The lower part shows the evidence of polaron hopping and MR behaviour at room temperature when a magnetic field is applied in a parallel direction [87].

### 6.1.2 Zinc Oxide (ZnO)

Although $TiO_2$ is the one where the first report of DMOs is observed, but, ZnO is the prominent candidate for where maximum DMOs investigation had been done so for, as shown in Fig. 23. ZnO is an interesting candidate due to the wide Optical band gap of 3.37 eV having a wurtzite structure. The oxide had a natural tendency to be *n*-type due to interstitial defects and oxygen vacancies.

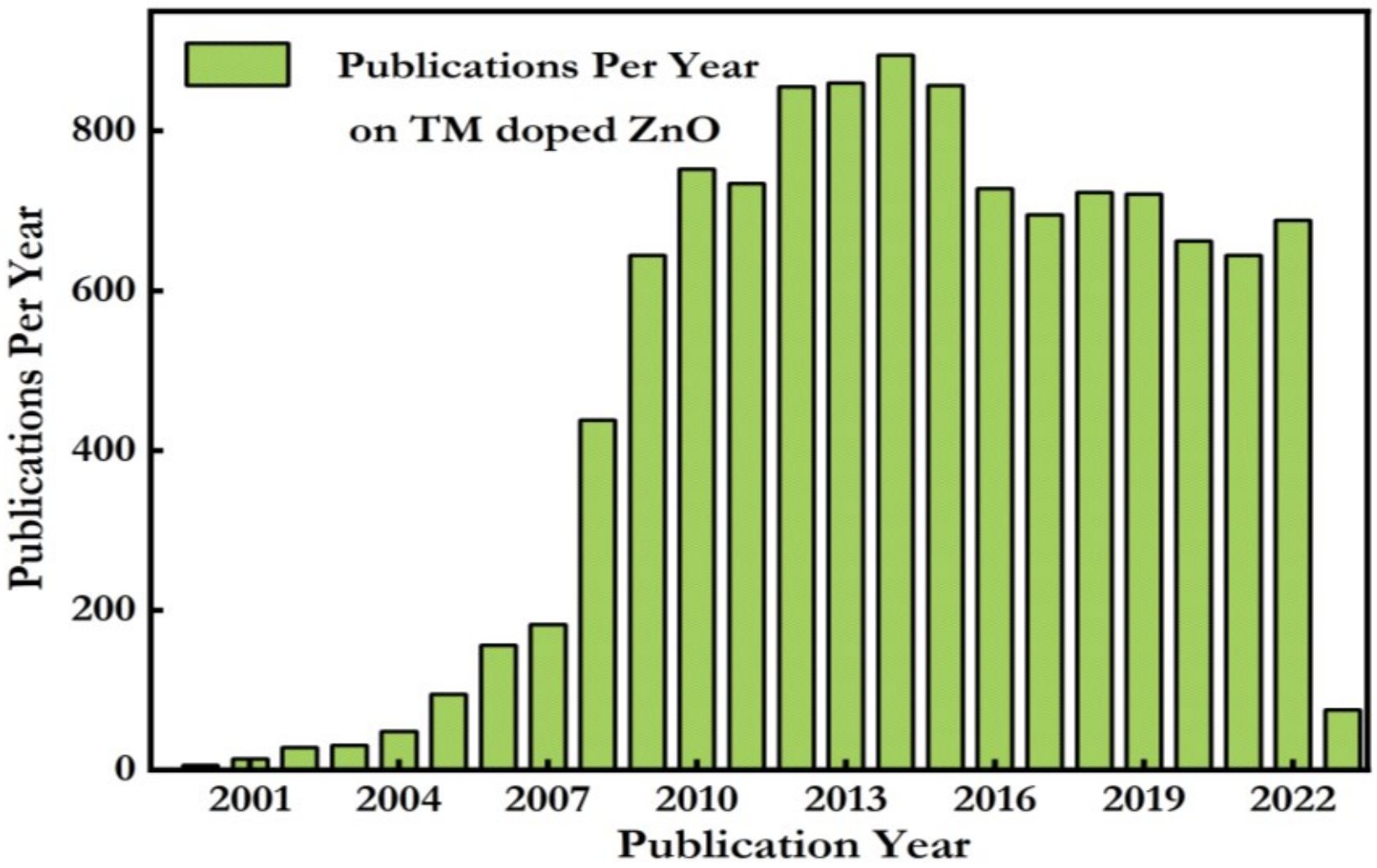

**Figure 23.** Number of publications per years on TM doped ZnO based on DMOs, according to Engineering Village database.

Ueda et al. reported first, in 2001, high $T_C$ in Co-doped ZnO thin films deposited by the PLD techniques up to 10% concentration and achieved saturation magnetization of 2.7 $\mu_b$/Co with *n*-type conductivity. The origin of magnetization is assumed to be indirect RKKY interaction and in higher Co doping secondary impurity [65]. In 2003, P. Sharma and colleagues et al. synthesized Mn-doped ZnO using a solid-state method of up to 4 % concentration by varying sintering temperature from 500°C to 900°C. They observed RTFM in sintering temperature above 700°C due to Mn cluster with a $T_C$ of 424 K. after then, they deposited Mn-doped ZnO thin films using PLD on quartz substrate sintered at 500°C and achieved RTFM in transparent thin films. The origin of RTFM was assumed in the theoretical estimation study done by Dietl et al. for Mn-doped ZnO [101]

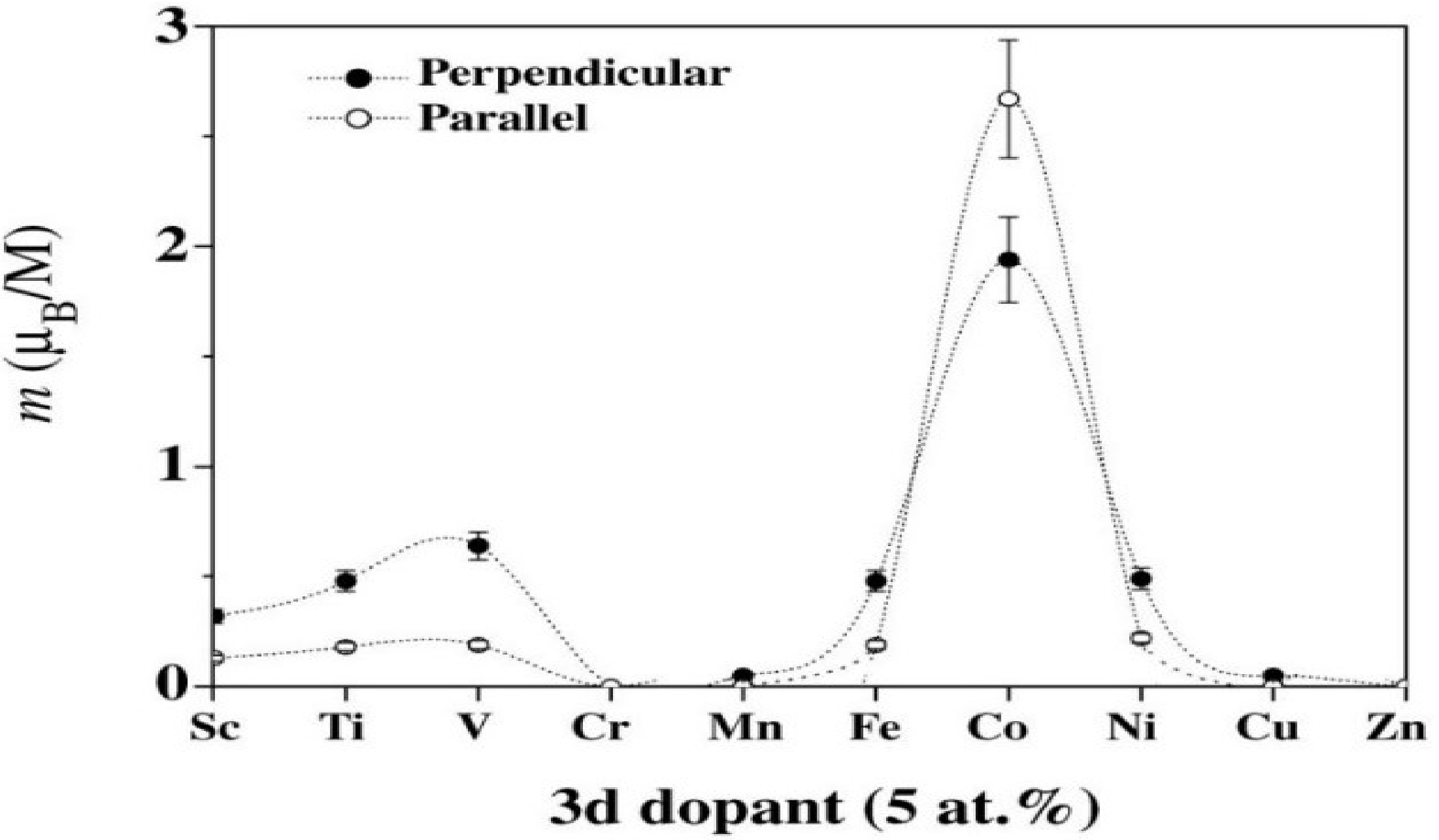

**Figure 24**. Magnetic moment of 3d dopant ions into ZnO thin films [102].

In 2004, M. Venkatesan and colleagues doped 5 % TM (Sc, Ti, V, Fe, Co, and Ni) ions in ZnO and deposited thin films on an r-cut sapphire substrate using PLD. They observed that magnetism is very anisotropic regarding the substrate and observed RTFM in all thin films, having the highest magnetic moment with Co-doped ZnO, as shown in Fig.24. They concluded that the observation of magnetic anisotropy is associated with the allied orbital around magnetic defects, which causes the origin of magnetic moment at room temperature [102]. K. Kittilstved (2006) and group deposited 0.2 % Mn and 3.5 % Co-doped ZnO thin films by direct chemical synthesis with or without the supply of nitrogen individually. The addition of nitrogen is attributed to the change in the conductivity of a thin film from *n*-type to *p*-type, which yields a better understanding of the origin of magnetism in DMOs. They observed RTFM with the optimal magnetic moment because of the p-d conductive hybridization of Mn and Co ions [103]. Z. Zhang and colleagues (2009) synthesized 5.6 % Co and 4 % Fe doped ZnO nanoparticles individually using solvothermal techniques and observed intrinsic FM at room temperature. To validate their evidence, they did HRTEM-EDS, ALCHEMI-EELS and EMCD measurements and concluded that the observed magnetism arises in Co-doped ZnO due to intrinsic FM, while with Fe-doped ZnO, it wasn't intrinsic [104]. R. Saleh et al. in 2012 synthesized Fe-doped ZnO nanoparticles using the co-precipitation method and achieved the solubility of Fe to 21 % with no secondary impurities. He observed the narrowing of the optical band gap and RTFM with saturation magnetization of 0.2629 emu/g. The observation of FM is attributed to the mixed valence state of Fe and the sp-d spin exchange interaction [105].

To find the origin of magnetism in undoped ZnO, T. Taniguchi and colleagues 2013 prepared a single sheet of crystalline ZnO using an electrodeposition technique of an approximate thickness of 1.5 nm. They observed higher $T_c$ than RT and an optical band gap of 3.5 eV. They concluded that the observation of magnetization is due to "*$d^0$ magnetization*" occurring due to a high surface-to-volume ratio [106]. Further, T. Tietze and colleagues in 2015 deposited two ZnO thin films one is coarse grain, and another one is fine grained using liquid ceramic methods. To find out the origin of FM, they employed the local probe method of low energy muon spin relaxation (LE–μSR) together with HRTEM and SQUID. They observed that the magnetization volume increases with the grain size reduction of the samples and that the internal magnetic field is independent of grain size distribution [107].

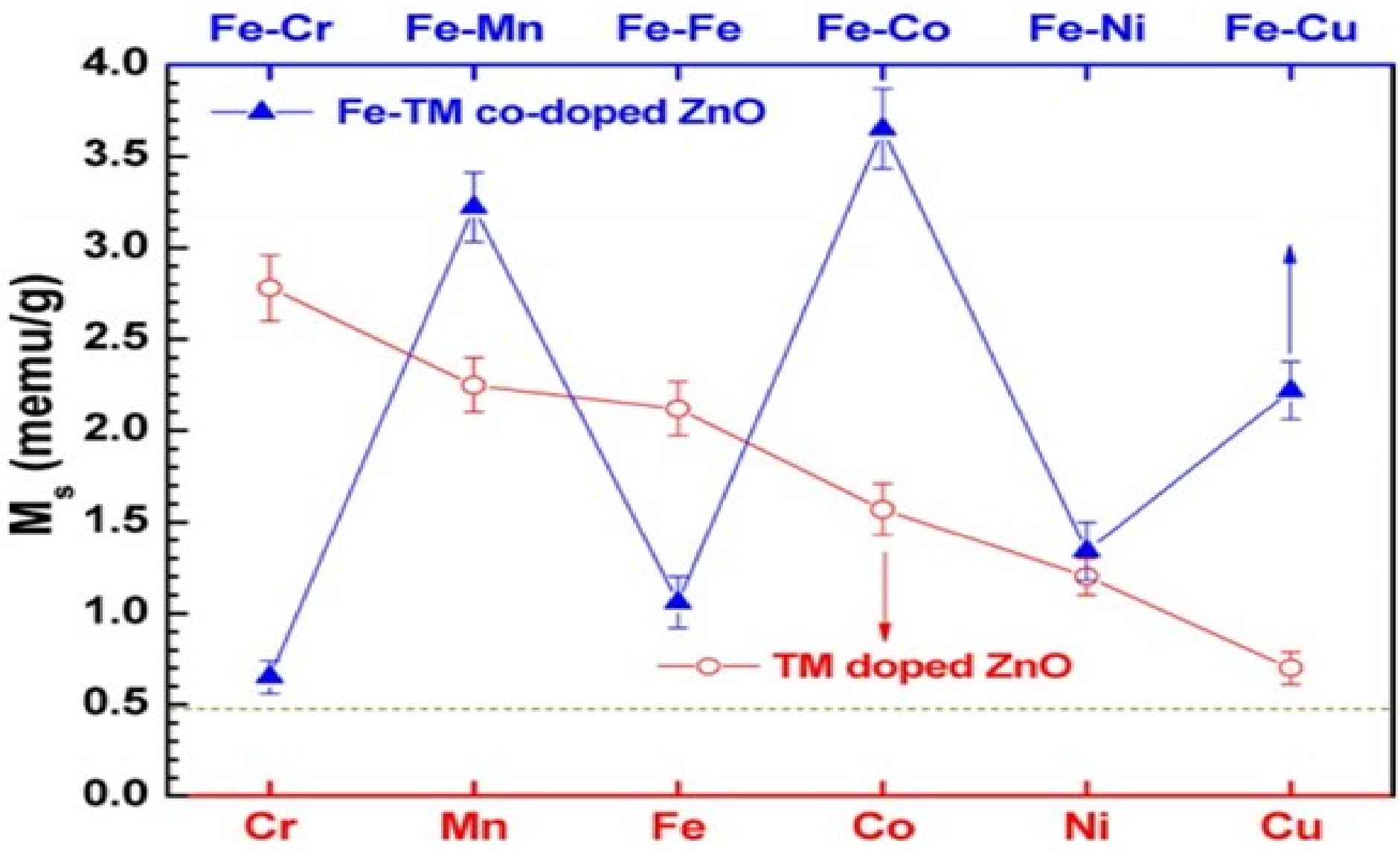

**Figure 25**. Saturation magnetization of doped TM ions and co-doped Fe-TM ions into ZnO nanoparticles [108].

To find out the combined effects of TM ions into ZnO samples, J. Beltran et al. in 2016 doped 0.01 of TM (Cr, Mn, Fe, Co, Ni, and Cu) ions and co-doped Fe-TM ions into ZnO NPs synthesized using the modified Pechini method. The observation of RTFM in TM doped ZnO NPs are attributed to the lattice distortion, while in the case of co-doped Fe-TM ions attributed to the mixed oxidation state of TM ions results in charge transfer mechanism and grain boundary effects, the saturation magnetization variation of TM ions and co-doped Fe-TM ions into ZnO NPs shown in Fig.25 [108].

To understand the mechanism of DMOs, in 2019, Nasir et al. Deposited non-magnetic $Cu^{2+}$ ions substituted ZnO thin films in quartz substrate using RF sputtering and observed intrinsic RTFM with the higher magnetic moment of 1.87 $\mu_b$/Cu (0.05), associated with the mixed valence states of Cu, results in the formation of defects and BMP [109]. Further, K. Poonia et al. in 2021 doped 4f element $Gd^{3+}$ ions into ZnO synthesized by the auto combustion sol-gel method up to 10 % concentrations and observed that the AFM dominated FM contributions associated with the intrinsic defects and uncompensated magnetic moment through ZnO lattice [110]. N.C. Ani and colleagues (2022) investigated the performance of organic spintronics by depositing Gd-substituted ZnO thin films using the spin-coating sol-gel method up to 10 % atomic concentrations. The optical studies revealed higher average transparency, while hall effect measurements showed anomoulu8us hall effect (AHE), and MFM measurement confirmed the intrinsic FM and spin polarization at room temperature. The MR studies show positive MR of 7.2% at room temperature, which was beneficial for spin injection for organic spintronics [111].

### 6.1.3 Indium Oxide ($In_2O_3$)

Indium oxide ($In_2O_3$) is a widely used transparent *n*-type transparent conducting oxide with high carrier concentrations and mobility, making it an exciting candidate for DMOs with the discovery of $In_2O_3$-based DMSs.

The first reported RTFM in Mn-doped $In_2O_3$ thin films by J. Philip et al. (2004) was grown on Si/$SiO_2$, Si, and sapphire substrate using reactive thermal evaporation. They observed high magnetization in the case of a sapphire substrate of 0.8 $\mu_b$/Mn with higher transparency and n-type conductivity having a carrier concentration of $2.5 \times 10^{19}$ $cm^{-3}$, the observation of magnetization attributed to indirect RKKY interaction [112]. A year later, in 2006, J. Philip and colleagues again grew 2 % of Cr in $In_2O_3$ by varying oxygen flow in Si/$SiO_2$ substrate using reactive thermal evaporation. They obtained a highly transparent, wide band gap with n-type conductivity. They also observed RTFM with $T_c$ of 890-913 K. They concluded that the observation of high $T_c$ was dependent on higher carrier density and observation of magnetization was dependent on the occurrence of magnetic polarons, which were controlled by the oxygen flow [112].

R. Panguluri (2009) and their group also studied the 2 % Cr doping on $In_2O_3$ thin film deposited on a sapphire substrate using a reactive evaporation method. They observed RTFM with a magnetic moment of 0.5 emu/$cm^3$ arises from the 50 % spin polarization occurs due to carrier-mediated interaction, which was confirmed from point contact Andreev reflection measurements [113]. T. Bora (2011) et al. synthesized Co-doped $In_2O_3$ nanoparticles up to x = 0.07 from the solid-state method. They observed $T_c$ of 718 K and a magnetic moment of 0.09 emu/g (x =0.07) due to the formation of BMP and a change in magnetic anisotropy [57]. S. Farvid (2014) group studied the effects of the charge transfer mechanism in Mn (2 to 20 %) doped $In_2O_3$ nanocrystals synthesized by colloidal nanocrystals and further spin-coated for thin films. XAS study revealed mixed valence states of Mn, and XMCD confirms the origin of FM order due to evidence of charge transfer mechanism [114]. Additionally, Sena (2015) et al. studied the effect of charge transfer with 5% TM (Fe, Ni, Co) ions doping into $In_2O_3$ NPs synthesized by Pechini sol-gel method and observed RTFM with a high magnetic moment of 0.8 $\mu_b$/Fe, in case of Fe doping. They also did first-principles calculations based on a density functional theory (DFT) study and experimental Perturbed angular correlation (PAC) spectroscopy to confirm the charge transfer phenomenon. They confirmed that oxygen vacancies play a vital role in charge transfer, thus, in enhanced magnetic properties [115].

L. Shen's (2017) group fabricated N (nitrogen) doped $In_2O_3$ thin films using RF magnetron sputtering on Si (100) and glass substrate. They observed that the conductivity $In_2O_3$ changes from *n*-type to *p*-type with N-doping because of p-p interactions between N-2p orbitals. The change in conductivity results in RTFM, an increase in higher Ms value, and the formation of positive MR at 10 K [116], as shown in Fig.26.

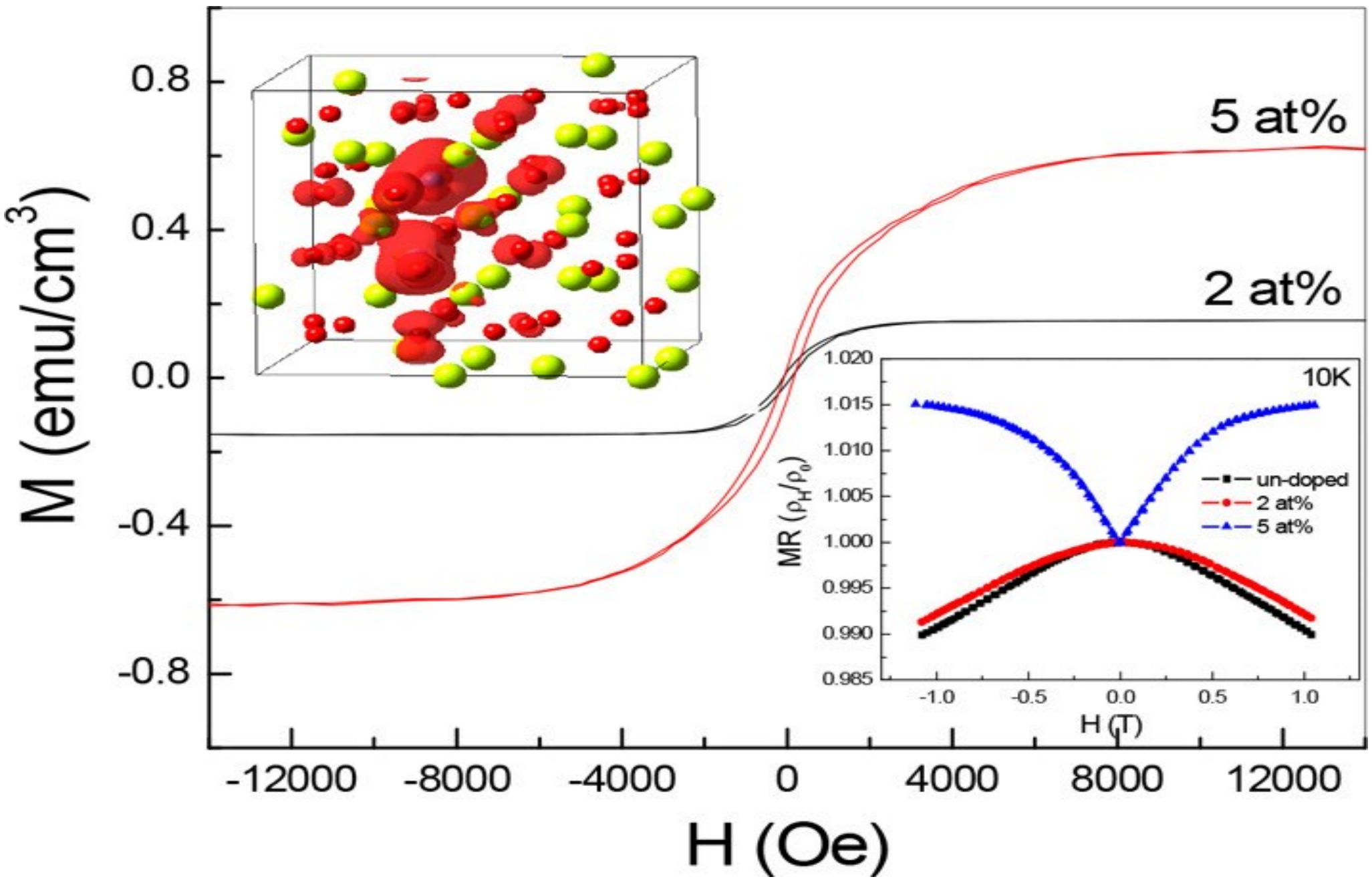

**Figure 26.** N-doping in $In_2O_3$ thin films shows higher Ms at 5 % concentrations at RT and positive MR at 10 K due to *p*-type conductivity [116].

Lou (2018) et al. deposited 5% of Fe into $In_2O_3$ thin film on a MgO substrate using PLD by varying oxygen flow. They observed RTFM in all thin films, and low energy muon spin relaxation measurement confirms the FM phase in the $In_2O_3$ lattice. The observation of magnetism arises due to the observation of oxygen vacancies resulting in the formation of BMP[95]. M. Dong (2022) et al. study the origin of magnetism in undoped $In_2O_3$ thin films prepared by magnetron sputtering by varying Ar/$O_2$ flow. They observed the formation of ordered porous $In_2O_3$ thin films showing RTFM having a magnetic moment of 18.78 emu/cm$^3$ due to the direct coupling of polarized electrons with oxygen vacancies [117]. Xu (2023) and group fabricated composite $In_2O_3$ thin films device of Ag/$In_2O_3$/PAA/Al heterostructures deposited by DC magnetron sputtering for magneto-resistive storage applications. They deposited thin films by varying sputtering pressure and power, which resulted in the observation of RTFM and the observation of high Ms of 42.03 emu/cm$^3$ with a sputtering pressure of 3.2 Pa because of originated oxygen vacancies. Further, memory switching devices exhibit substantial bipolar switching behaviour because of small Ag conducting filament, which results in dominant oxygen vacancies, which lays the foundation for $In_2O_3$-based memory storage devices for spintronics applications [118].

### 6.1.4 Cerium Oxide ($CeO_2$)

Cerium Oxide ($CeO_2$) is a diamagnetic insulating oxide, having a wide-band-gap of 3.4 eV and dielectric constant of 26 and is widely used as dilute magnetic oxide because of its data repeatability [119].

A. Tiwari (2006) reported the first $CeO_2$-based DMOs when they deposited 5% Co-doped $CeO_2$ thin films in an LAO substrate using the PLD technique and obtained highly transparent thin films. They also observe a significant magnetic moment of 8.4 $\mu_b$/Co atom with $T_c$ of 740-875 K. The XPS, EELS and HRTEM measurement confirms the observed magnetization is intrinsic and occurs due to the trapping of electrons in oxygen vacancies [120]. Later in 2007, B. Vodungbo et al. doped 4.5 ± 0.5 % Co epitaxial transparent $CeO_2$ thin films in STO and Si substrate using the PLD technique. They observed that the STO substrate had a higher magnetic moment of 1.7 $\mu_b$/Co atom at 400 K, ascribed to the strain-induced oxygen vacancies with Co doping [121]. Xiaobo et al. (2009) doped nonmagnetic Ca in $CeO_2$ NPs synthesized using solution combustion techniques up to 0.10 concentration without any secondary phase. They observed that all the samples exhibited FM nature at room temperature with diamagnetic contributions. The observation of magnetization is ascribed to the mixed valence state of Ce with Co doping occurs due to the observation of surface/ interface defects and oxygen vacancies [122]. Similarly, N. Paunovic (2012) et al. doped rare earth element Pr with 30% solubility without observation of secondary phase in $CeO_2$ NPs prepared by self-propagating room temperature synthesis method and observed decrement in the magnetic moment with Pr doping attributed to the occurrence of mixed oxidation states and oxygen defects mediated AFM superexchange interactions [123].

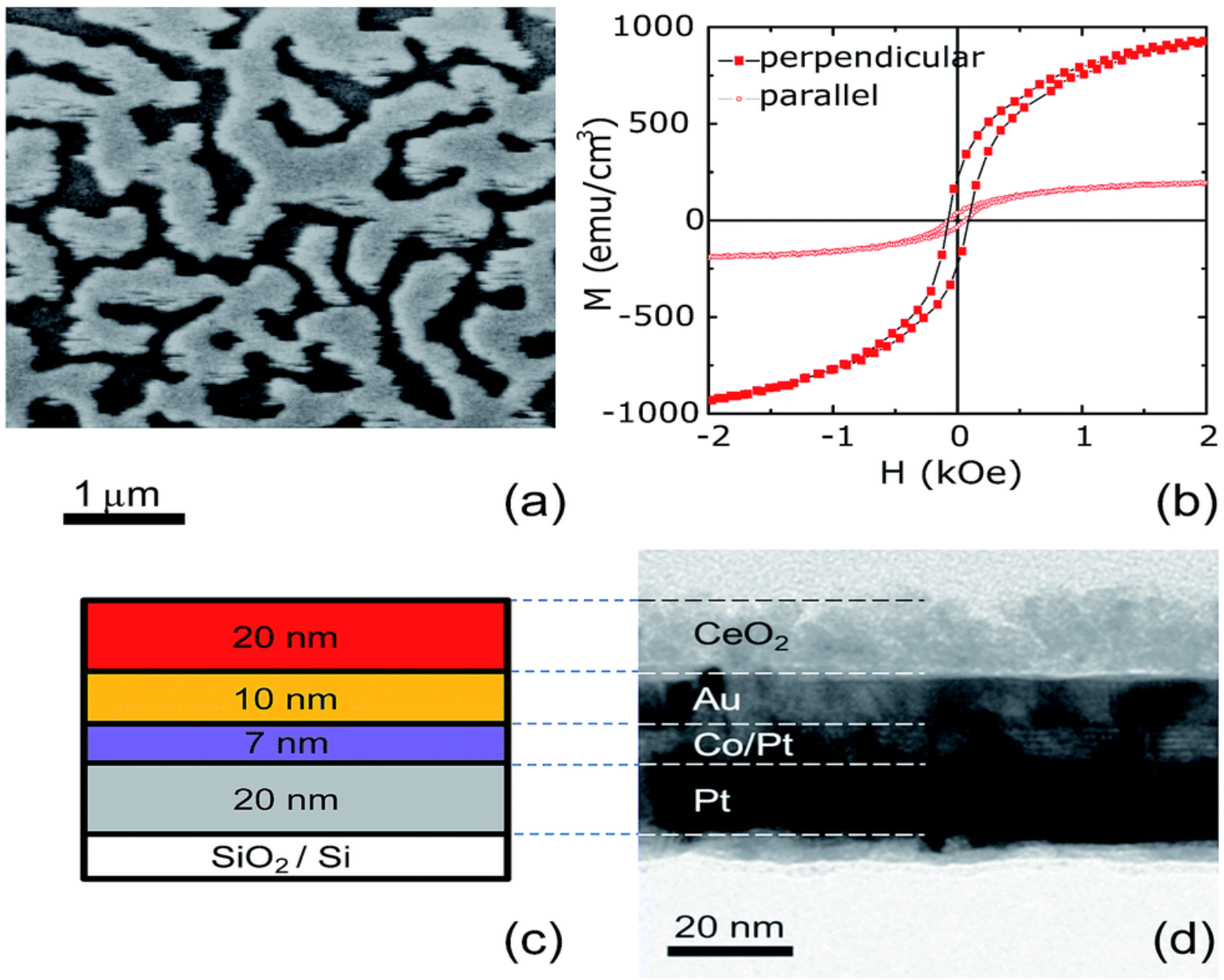

**Figure 27**. The multilayer heterostructure of $CeO_2$ fabricated by the Nicholas group shows perpendicular magnetic anisotropy because of the $CeO_2$ interfacial layer confirmed by TEM and MFM imaging [124].

K. Ranjith and colleagues (2014) prepared Co-doped $CeO_2$ NPs up to 6 % doping by co-precipitation method and observed RTFM with saturation magnetization of 0.028 emu/g. The XANES measurement confirms the observation of a mixed valence state of Ce, which is likely to accompany oxygen relocation with Co doping [125]. Coey and group (2016) synthesized La (x = 0.1) doped $CeO_2$ NPs by precursor method with surface treatment and achieved a crystallite size of 4 nm. They observed magnetization up to 380 K and explained the phenomenon of giant orbital paramagnetism, which was firstly reported in mesoscopic quasi-2D material [50]. Nicholas and group (2016) studied the effects of perpendicular magnetic anisotropy (PMA) in the Co/Pt multilayer heterostructure with $CeO_2$ fabricated using sputtering on Si (100) wafer, which was confirmed by MFM imaging and by HRTEM cross-sections as shown in Fig. 27. They observed that multilayer shows distinctive MH hysteresis at room temperature in respect to the applied magnetic field. The enhancement of hysteresis is observed when that perpendicular magnetic field is applied on the multilayer. The enhancement of hysteresis attributed to stable PMA occurs because of the interfacial $CeO_2$ layer. The PMA stabilization can benefit perpendicular MTJ for magnetic recording spintronics devices [124]. R. Murugan (2017) studied the doping of Mn with 15 % solubility in $CeO_2$ thin films deposited on a glass substrate using RF magnetron sputtering. They observed RTFM in all thin films due to originated defects such as oxygen vacancies [80].

### 6.1.5. Tin Oxide ($SnO_2$)

Tin Oxide ($SnO_2$) is a significant n-type transparent wide-bandgap of 3.4 eV semiconductor with high transparency and chemical stability, having potential applications in gas sensors and optoelectronics properties [126].

S. Ogale and group (2003) reported a first-time giant magnetic moment of 7.5 ± 0.5 $\mu_b$/Co atom in transparent Co-doped Rutile $SnO_2$ transparent thin films grown or r-cut sapphire substrate using PLD technique with $T_c$ close to 650 K. the observation of magnetic moment attributed to the unquenched orbital moment of Co ions [89]. C. Fitzgerald (2004) et al. doped Fe, Mn, and Co ions of 5% concentration in $SnO_2$ ceramic material synthesized by solid-state method and observed RTFM with $T_c$ of 360 K, without any secondary impurity. The observation of RTFM was concluded to FM exchange coupling in the $SnO_2$ lattice, which involves trapping electrons close to oxygen vacancies with nearer TM ions [127]. P. Archer (2006) et al. studied the role of grain boundary defect as they obtained $T_c$ of 350 K, with the 0.48 % $Ni^{2+}$ ions substitution in $SnO_2$ NCs prepared by direct chemical method. The observation of FM was attributed to dependence on the synthesis conditions, as the low annealing temperature caused NC to aggregate results in interfacial grain boundary defect, which was activated by oxygen vacancies [128]. C. Komen (2008) doped 0-12 % of TM (V, Cr, Mn, Fe, Co, and Ni) cations into $SnO_2$ NPs employed by the sol-gel method. They observed FM increases up to specific concentrations and afterwards decreased. They observed conclusive magnetic moment in Co, Fe and Cr dopants because of the structural distortion in the $SnO_2$ lattice with cations doping, which causes the creation of oxygen vacancies [129].

P. Wu (2012) observed RTFM with doping of 6 % nonmagnetic Mg ions into the transparent epitaxial $SnO_2$ thin film deposited by RF-magnetron sputtering. The observed magnetization of 6.9 emu/$cm^3$ having *p*-type conductivity occurs due to carrier-induced hole indirect exchange interactions and oxygen vacancies [130]. In the same year, Ghosh (2012) and colleagues deposited rare-earth Gd-doped $SnO_2$ thin films in up to 5% concentrations using RF-sputtering and observed that with Gd substitution, the conductivity of $SnO_2$ changes from *n*-type to *p*-type. It is also observed that pure $SnO_2$ shows RTFM, while with doping, the magnetization diminishes because of a reduction in oxygen vacancies as the resistivity of materials increases, which indicates that the *p*-type conductivity or magnetic dopant is not totally responsible for RTFM [131]. Wang (2014) et al. observed the $d^0$ FM in epitaxial Li-doped $SnO_2$ films grown on a c-cut sapphire substrate using RF-magnetron sputtering. They observed the change in conductivity from *n*-type to *p*-type with high transparency and magnetic moment of 7.9 emu/cm$^3$, ascribed to FM coupling exchange interactions between p-p orbital of oxygen and unpaired 2p electrons [74].

N. Salah (2016) and group doped Mn (x = 0 to 5 %) ions into $SnO_2$ NPs using the microwave-assisted method and observed the widening of the optical band gap and RTFM with Mn doping. They observed FM arises from grain boundary interfaces because of oxygen and Sn vacancies[132]. Kusuma (2018) et al. demonstrated the magnetic behaviour of 0 to 5 % doped $Cr^{3+}$ ions into $SnO_2$ lattice prepared by combustion method and observed RTFM in all NPs samples. The occurrence of RTFM were ascribed to the mixed valence state of Sn and Cr ions, which acts as F centres and trapped conduction electrons, and their interaction cause FM order [35].Xinran and group (2019) investigated the impact of resistive and electric switching through magnetic studies on Ta/$SnO_2$/Pt/Ti/$SiO_2$/Si and $SnO_2$/Pt/Ti/$SiO_2$/Si thin film devices using a sputtering technique by varying annealing temperature and Ar/$O_2$ flow. They observed RTFM in both devices with minimal change in saturation magnetization, confirming the minimal effect of the Ta electrode in magnetization, as shown in Fig.28. It is also observed that a low resistive state shows higher Ms than a high resistive state because of the formation of oxygen vacancies, which causes higher Ms. They observed that with the Ta electrode device, the had very high multilevel resistive switching and huge ON/OFF ratio of 2700, which is beneficial for high-density data storage applications[133].

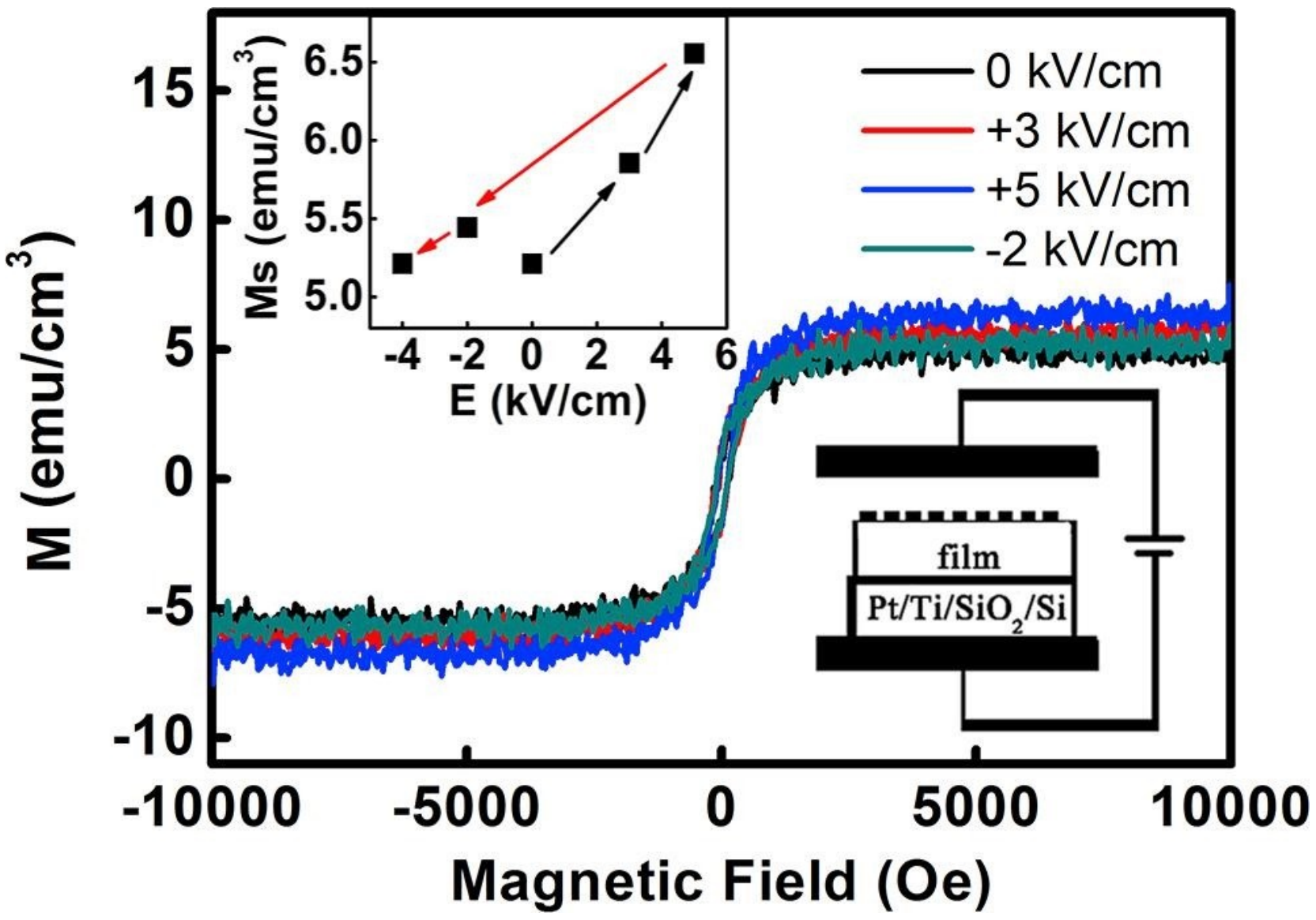

**Figure 28.** Ta/$SnO_2$/Pt/Ti/$SiO_2$/Si thin film device shows RTFM of with the effect of applied electric field [133].

D. Manikandan (2022) did a very comprehensive review of the FM behaviour of $SnO_2$ material, dependent on their size, morphology in undoped and doped $SnO_2$ bulk, nanostructure, and thin films with their magnetism models and correlated with the DFT studies. S.C. Ray (2022) demonstrates the single valence state Fluorine (F) doping on $SnO_2$ thin films prepared by CVD, studied their electronics and magnetic characteristics, and observed the higher density of states in F-doped $SnO_2$ thin film, which results in higher electrical conductivity and RTFM [134] S. Pat and colleagues investigated the effect of Fe, Co, and Fe-Co co-doped $SnO_2$ thin films grown by thermionic arc deposition system on various substrates and studied their structural, optical, and magnetic behaviour. Their structural studies confirm the observation of secondary phases in Fe and Co-doped samples while a single-phase Fe-Co doped $SnO_2$ thin films. They observed high transparency and wide optical band gaps with RTFM in all thin films. It is also observed that the substrate's nature also influences the magnetic characteristics of grown thin films because of material microstructure and crystal imperfections [135].

### 6.2 *p*-type based DMOs.

In today's scenario, every oxide that ever existed, mainly *n*-type such as ZnO, $TiO_2$, $SnO_2$, $In_2O_3$, etc., doped with every TM ion to become the best candidate for DMOs, as discussed in previous literature sections. Originally Dietl et al. theoretically predicted that FM existed only in hole-rich regions or in *p*-type DMOs, which makes the researcher convert conductions from *n* to *p*-type by TM ions substitutions in materials or by changing the synthesis parameters as mentioned earlier in oxides such as ZnO, $SnO_2$ and $TiO_2$.

The *p*-type oxides as a DMO candidate are an unexplored area till now, with minimal studies being done so far. Recently, *p*-type oxides have garnered interest in the semiconductor industry because of their high application base as optoelectronics devices, which can be combined with magnetic TM ions for potential spintronics device applications as the subsequent DMO material candidates.

#### 6.2.1 Nickel Oxide (NiO)

Nickel Oxide (NiO) is an antiferromagnetic Mott-Hubbard insulator with $T_N$ lying around 523 K and a wide optical band gap of 4 eV having *p*-type semiconducting properties.

J. Wang and colleagues (2005) first reported first NiO-based DMO system nanoparticles with Fe doping up to 0.02 concentration prepared using-precipitation method and observed RTFM. They attributed that the observation of high Ms is not entirely attributed to finite size effects and concluded that the observed RTFM is because of FM cluster [136]. Since the observation of the first NiO-based DMS, there have been minimal studies because of low conductivity and origin. S. Layek (2015) et al. observed the phenomenon of RTFM with Mn doping ($0 \leq x \leq 0.06$) in NiO NPs synthesized by low-temperature hydrothermal method. They observed very high $T_c$ of about 653 K (0.02 Mn), then afterwards decreased, the observation of high magnetization and $T_c$ attributed to double exchange interaction between Mn-Mn atoms and the canting of AFM lattices [54]. Bharathy. G (2017) et al. studied pseudo capacitance and magnetic properties of Co-doped NiO NPs synthesized using the sol-gel method. They observed high saturation magnetization with Co doping attributed to the secondary phase and uncompensated surface spins [137].

M. Siddique (2020) and their group studied optical and magnetic behaviour with the effects of $Ce^{3+}$ ions doping in NiO lattice prepared using the solgel method. They observed the narrowing of the optical band gap, ascribed to lattice disorder with Ce ions doping. They also observed that with Ce doping there is an increase in magnetic moment attributed to the mixed valence state of Ce, confirmed by XPS analysis and further EPR analysis corroborate the existence of unpaired electron results in the formation of oxygen defects [56]. A. Chhaganlal and group (2021) doped 4f element Sm ($x = 0.1 - 0.5$) into NiO NPs prepared by coprecipitation method and observed RTFM attributed to BMP model mediated by oxygen vacancies confirmed by XAS analysis [138]. D. Kaya and colleagues (2021) investigated various *p*-type characteristics and magnetic behavior of NiO thin films grown on multiple substrates, such as glass, sapphire, GaAs, InP, and Si, for better characteristics. They observed the diffusion of Ni ion with GaAs, InP, and sapphire substrate, resulting in secondary phase impurities, as their interaction is shown in Fig. 29. While sapphire and glass show high transparency and wide optical bandgap with *p*-type conductivity, and additionally show RTFM behavior. They observed higher Ms and Hc with glass substrate than sapphire because of deposition parameters caused by the cationic $Ni^{2+}$ or oxygen vacancies [139].

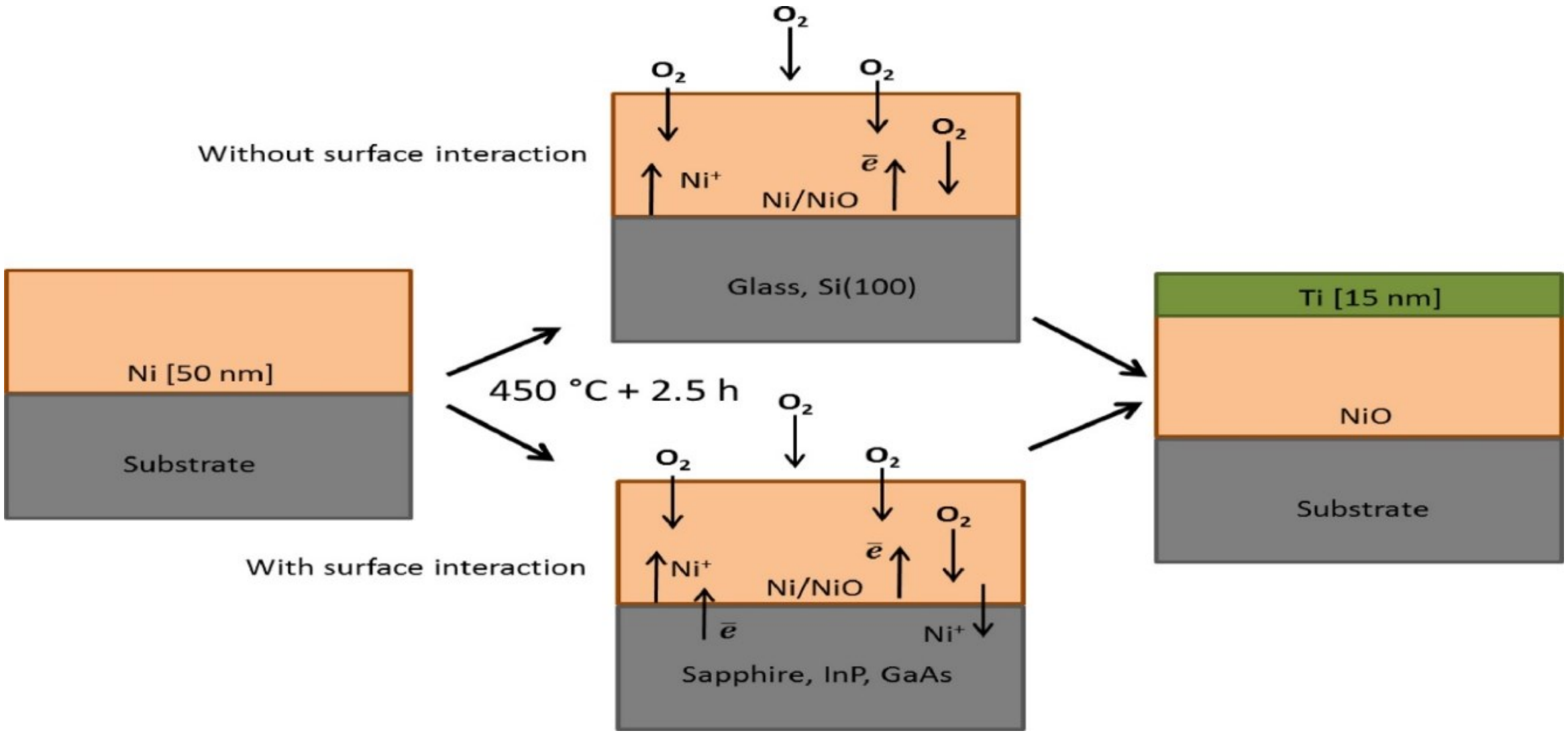

**Figure 29.** Interaction of NiO with various substrates [139].

The insulating behaviors of NiO and low carrier concentration limits the possibility of the origin of FM, which restricts NiO from *p*-type DMOs candidate.

### 6.2.2 Cupric Oxide (NiO)

Cupric Oxide (CuO) is an AFM p-type semiconductor with an optical band gap of 2.2 eV and $T_n$ of 212-230 K, which limits the possibility of the exposition of CuO as a DMOs candidate. However, numerous researchers observe RTFM in doped and undoped CuO about various origins of DMOs.

S.G Yang (2003) and colleagues first explored the possibility of FM in CuO nanoparticles by Mn doping from 3.5 – 15 at% prepared using the co-precipitation method. They observed that the FM transition at 80 K, confirmed by the C-W law, occurs due to the metal-insulator transition. They also observe positive MR below transition and negative MR above transition. They concluded that positive MR and high magnetization below transition were observed because of effective electron-phonon interactions and Zener exchange interactions [140]. Later, H. Zhu and the group (2006) deposited Mn-doped CuO thin films on oxidized Si substrate using RF magnetron sputtering from 6.6 to 29.8 % concentration. They observe FM transitions from 87-100K because of FM and AFM super-exchange interactions between Mn and Cu ions [141]. H. Qin et al. (2010) observed RTFM in CuO nanoparticles and thin films prepared by the sol-gel method because of the surface effects and observations of oxygen vacancies[142]

D.P. Joseph (2012) and group investigated the effect of Fe alloying on CuO thin films grown using spraying in the glass substrate and observed RTFM with a wide coercivity of hard magnets having *p*-type semiconducting behaviour. The occurrence of hard magnets is attributed to the occurrence of the mixed valence state of Fe and pinning of the film layer [143]. Samar Layek (2013) et al. explored the possibility of Fe (x = 0 to 0.06) doped CuO NPs prepared by the low-temperature sol-gel method. They observed that FM behaviour greatly enhanced with Fe doping up to x = 0.04 concentrations with $T_c$ of 350 K. The origin of FM attributed to structural disorder with Fe doping and magnetic anisotropy [144] .

K.G. Yang et al. (2016)., grew CuO epitaxial thin films on MgO substrate using plasma-assisted MBE and observed RTFM attributed to the mixed valence states of Cu ions and incurred oxygen vacancies, as observed with $d_0$ magnetization [145]. E. Batsaikhan (2020) and their group studied the finite-size effects of bare CuO NPs on magnetic behaviour. They observed that enhancement of FM property in CuO NPs dependent on the average crystallite size, as it felicitates charge redistribution triggered by lattice imperfection [146].

### 6.2.3 Chromium Oxide ($Cr_2O_3$)

Chromium oxide ($Cr_2O_3$) is a transparent conducting oxide (TCO) potential candidate owing to its wide optical bandgap of 3.1 eV, which had an application in photonic devices [147], light-based catalysts [148], electronics devices[149] , etc. $Cr_2O_3$ is an AFM material with a above room temperature, $T_n$ of 308 K. $Cr_2O_3$ is the first compound in which linear Magnetoelectric (ME) effects were theoretically predicted by D.N. Astrov et. al (1959) [150], experimentally observed, and fully exploited [151]. After well-studied phenomena of the ME effect in $Cr_2O_3$, recently, there has been a resurgence of interest in this field with the goal of use in technological applications like semiconductor spintronics, making it a viable candidate for DMO [152].

W.S. Zhang (2005) and colleagues investigated the structural and magnetic characteristics of Cr-coated $Cr_2O_3$ nanoparticles and observed AFM characteristics of Cr and $Cr_2O_3$ at the higher field and weak RTFM at the lower field. They attributed that the observation of RTFM might be because of exchange interactions between the uncompensated surface spins [153]. E. Winkler et al. (2006) synthesized Co-doped $Cr_2O_3$ nanoparticles using the solid-state reaction of up to 10% concentration and observed mixed FM and AFM states at room temperature. They attributed the observation of mixed phases to FM and AFM surface exchange interactions because of exchange biasing [154]. Sahoo et al. (2007) fabricated the $Cr_2O_3$ epitaxial thin film on a sapphire substrate using MBE and observed stress-induced magnetization termed piezo-magnetism. They observed a significant reduction in $T_n$ to 36 K from 307 K because of induced stress from pinning AF order results in exchange biasing. [155]. Y. Shiratsuchi. et al. (2009) investigated the effect of magnetic Co interface of 1nm thickness with AFM $Cr_2O_3$ on Au, $Cr_2O_3$, and Si substrates using MBE. They observed the RTFM because of the exchange coupling of Cr and Co spin, which causes the exchange biasing [156], as shown in Fig. 30. S. Punugupati et al. (2014,2015) fabricated epitaxial $Cr_2O_3$ thin films in various substrates, such as cubic yttria-stabilized zirconia coated Si (100), and r-cut sapphire grew using pulsed laser deposition (PLD). They observed RTFM in all grown thin films with a $T_c$ of above 400 K having a high saturation magnetization and coercive field. They concluded that lattice strain caused in the thin films results in higher oxygen defects, culminating in RTFM [157][152]. Carey et al. (2016) theoretically investigated the substitution of low valence transition metal cations into the $Cr_2O_3$ lattice, which could be considered an excellent candidate in the field of DMOs as low valence doping on lattice causes disproportion of charges, which can result in structural disorder and defects, hence higher the optical and magnetic characteristics in the oxide lattice [158].

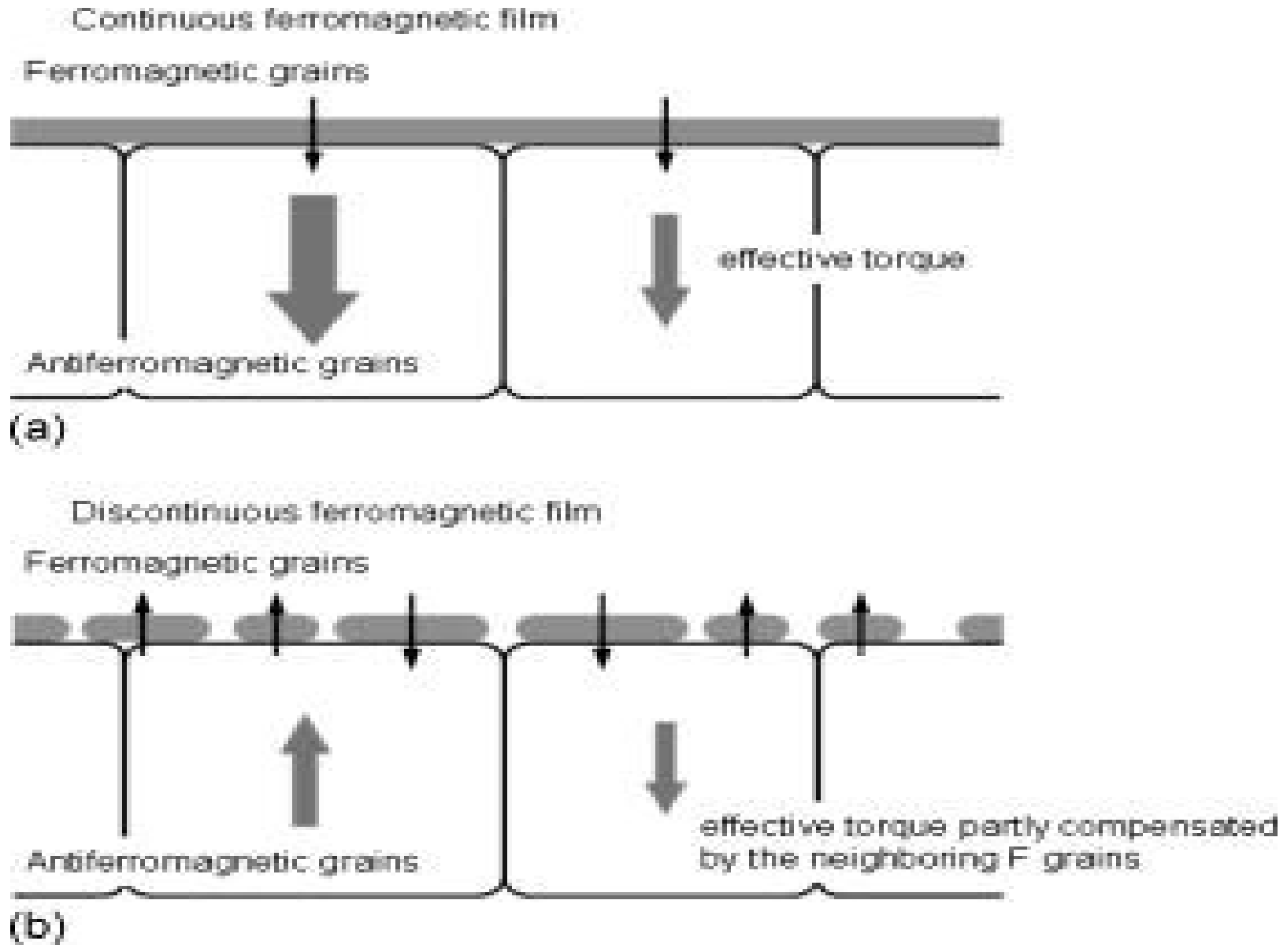

**Figure 30.** The interaction of the moment of magnetic Co ions with $Cr_2O_3$ (shown by the arrow), which produces torque from Co, depends on the thin film's surface structure [156]

Z. Qiu and colleagues (2018) investigated the phenomenon of spin-colossal magnetoresistance (SCMR) in trilayer YIG/ $Cr_2O_3$ /Pt devices fabricated on $Gd_3Ga_5O_{12}$ substrate grown using liquid phase epitaxy method. They inject spin current through the YIG layer into the $Cr_2O_3$ lattice layer for the spin current collection because of spin-hall effects, and the current is further transmitted to the heavy metal Pt layer. They observed the occurrence of spin-conduction and non-conduction transition around 296 K (below $Cr_2O_3$ $T_n$) because of the modulation of varying magnetic fields and termed the phenomenon SCMR. The observation of SCMR is attributed to the geometry of trilayer devices, which influences the Neel vector and magnetic anisotropy [159]. P. Bharaskar et al. (2019) investigated the influence of 2% Ti in $Cr_2O_3$ lattice thin films grown using PLD on a quartz substrate. They observed weak RTFM with high magnetic saturation and noticeable coercivity. They attributed that the observation of RTFM might be associated because of the Ti ions influence, which results in the occurrence of oxygen vacancies [160].

P. Makushko and colleagues (2022) discovered the phenomenon of flexomagnetic in $Cr_2O_3$ thin film growth using RF magnetron sputtering on the sapphire substrate, as in the form of Pt/ $Cr_2O_3$/ $Al_2O_3$ trilayer device. They discovered that the slight gradient of mechanical strain influences the $Cr_2O_3$ magnetic properties and $T_n$. They observed the magnetic moment per unit strain-induced gradient of 15 μb/nm$^2$ at the optimum thickness of 50 nm. They combined magneto transport and NV magnetometry on the grown device. They observed transition temperature at 100$^{\circ}$C above $Cr_2O_3$ transition with no degradation up to 1.5 years because of oxide epitaxy, as shown in Fig. 31. They anticipated that the origin of flexo magnetism caused by the strain induction, which results in the uncompensated spins at the top and bottom layer of the thin films [161].

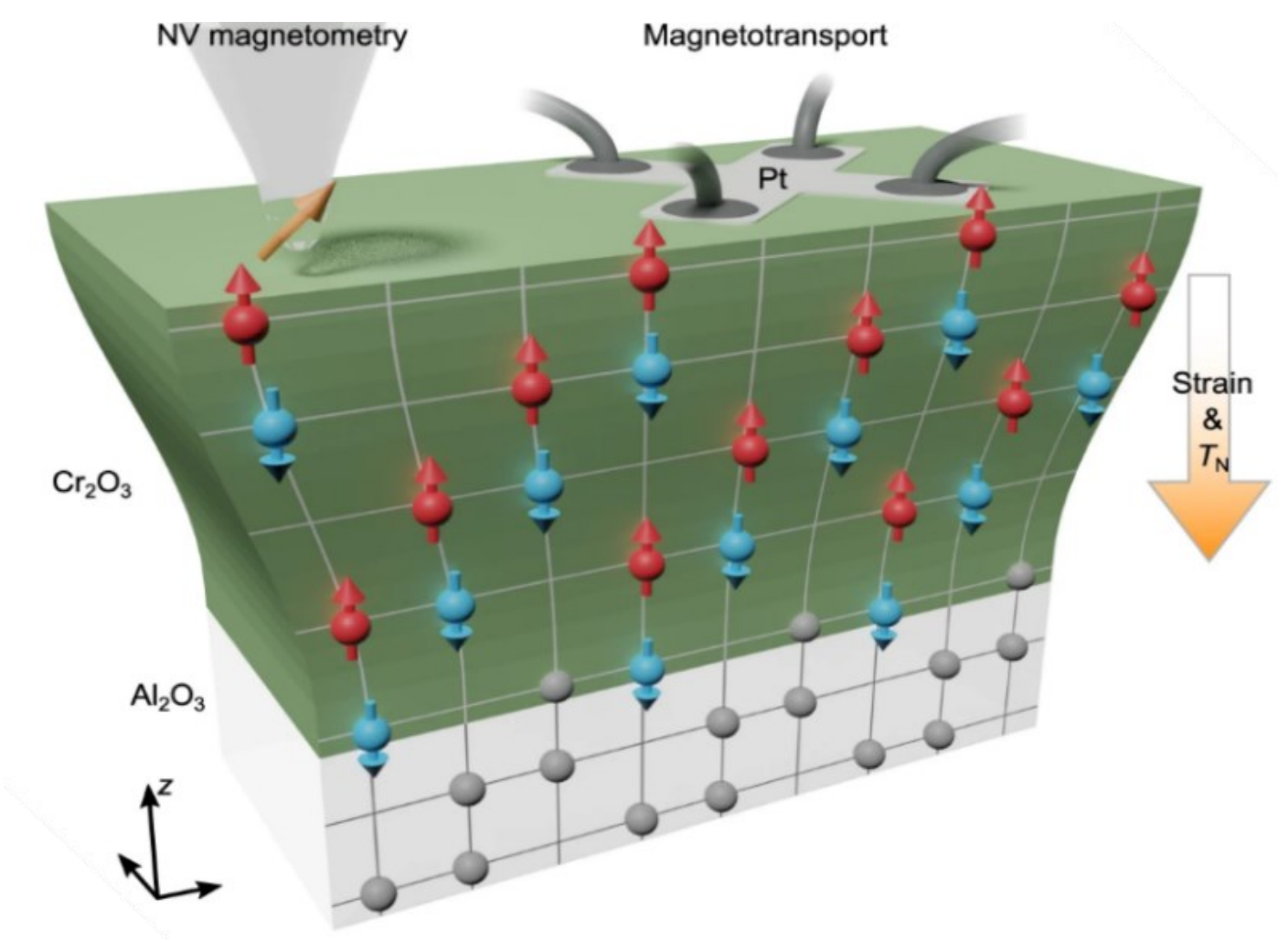

**Figure 31.** The effect of uncompensated spins in $Cr_2O_3$ trilayer device surface because of induced strain [161].

In this regard, our research group (2021-2024) extensively investigated the exposition of chromium oxides as an oxide based DMS. We investigated origin and control of magnetism in pristine $Cr_2O_3$ and TM (Ni, Fe, Mn, and Co) ions substituted $Cr_2O_3$ nanoparticles and thin films. We prepared TM ions substituted $Cr_2O_3$ nanoparticles up to 0.10 concentrations using the co-precipitation method without any secondary impurities and observed narrowing of optical band gap and RTFM in case of Ni substitution, while AFM nature in case of Co and Mn substitution and mixed FM and AFM at 300 K in case of Fe substitutions. The observation of various magnetic phases at room temperature was attributed to the observation of mixed-oxidations states and originated oxygen defects, which resulted in the formation of multiple super-exchange interactions and bound magnetic polarons model. We also grew the epitaxial $Cr_2O_3$ thin films with 5 % Ni, Mn, and Co concentrations on the *c*-cut sapphire substrate using PLD techniques. We observed high transparency with *p*-type conductivity having RTFM with a high $T_c$ of above 330 K in all grown thin films. We proposed a dual model for the observation of magnetism at room temperature and 5 K attributed to indirect hole mediated RKKY interaction and BMP model, respectively. Our investigative study explored the optical, transport, and magnetic characteristics of $Cr_2O_3$ with or without TM doping. This can open paths for next-generation *p*-type TCO-based ferromagnetic semiconductors as the DMOs candidate in potential spintronics devices [88].

## 7. Conclusions and Outlooks

In this review analysis, we investigated the two types of system as predicted by Dietl et al., *n*-type and *p*-type based DMO systems, which exhibited the RTFM with semiconducting properties. However, the origin of magnetism in maximum literature was attributed to secondary phases of TM ions or oxides, closing of ions, induced defects, charge carriers, induced strains, and structural irregularities, which can be influenced by the synthesis or fabrications of oxides, type of dopant and their dosage concentrations. The types of conflicted theories and mechanisms for the origin of RTFM make it impossible to find the proper origin of DMS, as various factors also influence its reproducibility and thermodynamic stability.

Based on the literature analysis from the series of research publications based on the DMOs, following conclusions were drawn:

*a)* It was observed that the concentrations of TM ions can influence the characteristics of oxides. The low percolation threshold of TM ions makes the DMO quite impossible, with favorable FM and semiconducting characteristics for real-life applications, as higher concentrations lead to secondary impurities and clustering. Very little literature focuses on the atomic and chemical states of dopant concentrations. HRTEM, XPS, and XAS characterizing techniques can eliminate the possibility of the assumptions of multiple phases and magnetic clusters and make the role of dopant ions more precise to understand.

*b)* The reproducibility and longevity of the observed RTFM in oxides are still significant drawbacks in DMOs. This reproducibility majorly depends upon the synthesis of nanoparticles and thin film deposition parameters, which can influence the morphology and structural integrity of the oxides, resulting in varying outcomes. The worldwide standard synthesis conditions and parameters can easily overcome the hurdles to become a definitive DMO material.

*c)* In some literature, it was observed that the doping of the same TM ions in oxides resulted in varying mechanisms from the effects of charge carrier to intrinsic defects, and sometimes some oxides show the phenomenon of $d_0$ magnetization without supporting their claims theoretically or experimentally, which results in ambiguity in the control and theory of DMOs. The significant factors in the varying origins are the absence of the proper characterization methods and the absence of experimental and theoretical studies such as density functional theory (DFT), magnetic force microscopy (AFM), anomalous hall effect (AHE), and optical-Kerr microscopy in numerous works of literature, which can eliminate the assumed preconceived mechanism in the system and moves towards a universal mechanism of DMOs.

*d)* In the vast majority of reported literature regarding magnetic oxide-based semiconductors, it is observed that the studies of p-type DMO/DMS systems could be more extensive in literature, as also observed in our literature review analysis. Dietl's theoretical studies suggested that p-type oxides are ideal candidates for a DMO system. However, difficulties in the synthesis and fabrication and lower conductivity of oxides might be the reason for limited studies. The stability of oxide for the DMOs system is still a significant hurdle because of non-repeatability, which can be overcome by experiment standardization.

*e)* Over 12000 literatures reported the phenomenon of DMS/DMOs in the last two decades, however, hardly or very limited studies regarding the use of DMOs as an active layer in devices fabrications. The use of DMOs system for devices fabrications can be potentially used if it obeys the principles intrinsic nature of DMOs with repeatability, which at current stage is quite challenging.

The following outlook for the dilute magnetic semiconductors and oxide systems is to design a standard novel framework for experimentation and characterization to overcome the existing system's traditional drawbacks (reproducibility and control), which can be beneficial for commercial spintronics device fabrication. The DMO system can unlock the potential of spintronics devices, as discussed in the review and analysis, because of its multiple functionalities in various applications. The next outlook can be searching for the next exotic materials for DMO, two-dimensional (2D) magnetic semiconductors, which are still in the infant stage but theoretically predict thermodynamic stability with high curie temperature.

This comprehensive review summarized the brief history of spintronics to oxide-based semiconductors. This review gives more profound insight into the control and origin of DMOs, and their challenges based on their conductivities with potential DMO-based spintronics devices. Hence, more effort should be made regarding the longevity and control of DMOs, which can eventually lead to next generation spintronics devices.